\documentclass[twocolumn,showpacs,preprintnumbers,amsmath,amssymb,superscriptaddress, bibnotes]{revtex4}

\usepackage{graphicx}
\usepackage{dcolumn}
\usepackage{bm}
\usepackage{xspace}

\newcommand{\EF}{$E_\mathrm{F}$\xspace}
\newcommand{\Uf}{$\mathrm{U}~5f$\xspace}
\newcommand{\orb}[2]{$\mathrm{ #1 } ~ #2 $\xspace}
\newcommand{\hn}[1]{$h\nu #1~\mathrm{eV}$\xspace}
\newcommand{\EB}[1]{$E_{\mathrm{B}} #1~\mathrm{eV}$\xspace}
\newcommand{\pnt}[1]{$\mathrm{#1}$\xspace}
\newcommand{\Gm}{$\mathrm{\Gamma}$\xspace}
\newcommand{\lineXX}[2]{$\mathrm{#1} - \mathrm{#2} $\xspace}
\newcommand{\lineXXX}[3]{$\mathrm{#1} - \mathrm{#2} - \mathrm{#3} $\xspace}
\newcommand{\TN}[1]{$T_\mathrm{N}=#1~\mathrm{K}$\xspace}
\newcommand{\gm}[1]{$\gamma = #1~\mathrm{mJ/mol K^2}$\xspace}

\newcommand{\etal}{\textit{et al.}\xspace}

\newcommand{\UAl}{$\mathrm{UAl}_3$\xspace}
\newcommand{\UGa}{$\mathrm{UGa}_3$\xspace}
\newcommand{\UIn}{$\mathrm{UIn}_3$\xspace}
\newcommand{\UAlGaIn}{$\mathrm{U}X_3$ ($X=\mathrm{Al}$, $\mathrm{Ga}$, and $\mathrm{In}$)\xspace}

\begin{document}
\draft
\preprint{HEP/123-qed}

\title{Electronic structures of $\mathrm{U}X_3$ ($X=\mathrm{Al}$, $\mathrm{Ga}$, and $\mathrm{In}$) studied by photoelectron spectroscopy}

\author{Shin-ichi~Fujimori}
\affiliation{Materials Sciences Research Center, Japan Atomic Energy Agency, Sayo, Hyogo 679-5148, Japan}

\author{Masaaki~Kobata}
\affiliation{Materials Sciences Research Center, Japan Atomic Energy Agency, Sayo, Hyogo 679-5148, Japan}

\author{Yukiharu~Takeda}
\affiliation{Materials Sciences Research Center, Japan Atomic Energy Agency, Sayo, Hyogo 679-5148, Japan}

\author{Tetsuo~Okane}
\affiliation{Materials Sciences Research Center, Japan Atomic Energy Agency, Sayo, Hyogo 679-5148, Japan}

\author{Yuji~Saitoh}
\affiliation{Materials Sciences Research Center, Japan Atomic Energy Agency, Sayo, Hyogo 679-5148, Japan}

\author{Atsushi~Fujimori}
\affiliation{Materials Sciences Research Center, Japan Atomic Energy Agency, Sayo, Hyogo 679-5148, Japan}
\affiliation{Department of Physics, University of Tokyo, Hongo, Tokyo 113-0033, Japan}

\author{Hiroshi~Yamagami}
\affiliation{Materials Sciences Research Center, Japan Atomic Energy Agency, Sayo, Hyogo 679-5148, Japan}
\affiliation{Department of Physics, Faculty of Science, Kyoto Sangyo University, Kyoto 603-8555, Japan}

\author{Yoshinori~Haga}
\affiliation{Advanced Science Research Center, Japan Atomic Energy Agency, Tokai, Ibaraki 319-1195, Japan}

\author{Etsuji~Yamamoto}
\affiliation{Advanced Science Research Center, Japan Atomic Energy Agency, Tokai, Ibaraki 319-1195, Japan}

\author{Yoshichika~\=Onuki}
\affiliation{Advanced Science Research Center, Japan Atomic Energy Agency, Tokai, Ibaraki 319-1195, Japan}
\affiliation{Faculty of Science, University of the Ryukyus, Nishihara, Okinawa 903-0213, Japan}

\date{\today}

\begin{abstract}
The electronic structures of \UAlGaIn were studied by photoelectron spectroscopy to understand the the relationship between their electronic structures and magnetic properties.
The band structures and Fermi surfaces of \UAl and \UGa were revealed experimentally by angle-resolved photoelectron spectroscopy (ARPES), and they were compared with the result of band-structure calculations.
The topologies of the Fermi surfaces and the band structures of \UAl and \UGa were explained reasonably well by the calculation, although bands near the Fermi level (\EF) were renormalized owing to the finite electron correlation effect.
The topologies of the Fermi surfaces of \UAl and \UGa are very similar to each other, except for some minor differences.
Such minor differences in their Fermi surface or electron correlation effect might take an essential role in their different magnetic properties.
No significant changes were observed between the ARPES spectra of \UGa in the paramagnetic and antiferromagnetic phases, suggesting that \UGa is an itinerant weak antiferromagnet.
The effect of chemical pressure on the electronic structures of $\mathrm{U}X_3$ compounds was also studied by utilizing the smaller lattice constants of \UAl and \UGa than that of \UIn.
The valence band spectrum of \UIn is accompanied by a satellite-like structure on the high-binding-energy side.
The core-level spectrum of \UIn is also qualitatively different from those of \UAl and \UGa.
These findings suggest that the \Uf states in \UIn are more localized than those in \UAl and \UGa.
\end{abstract}

\maketitle
\narrowtext
\section{INTRODUCTION}
\begin{figure}
	\includegraphics[scale=0.5]{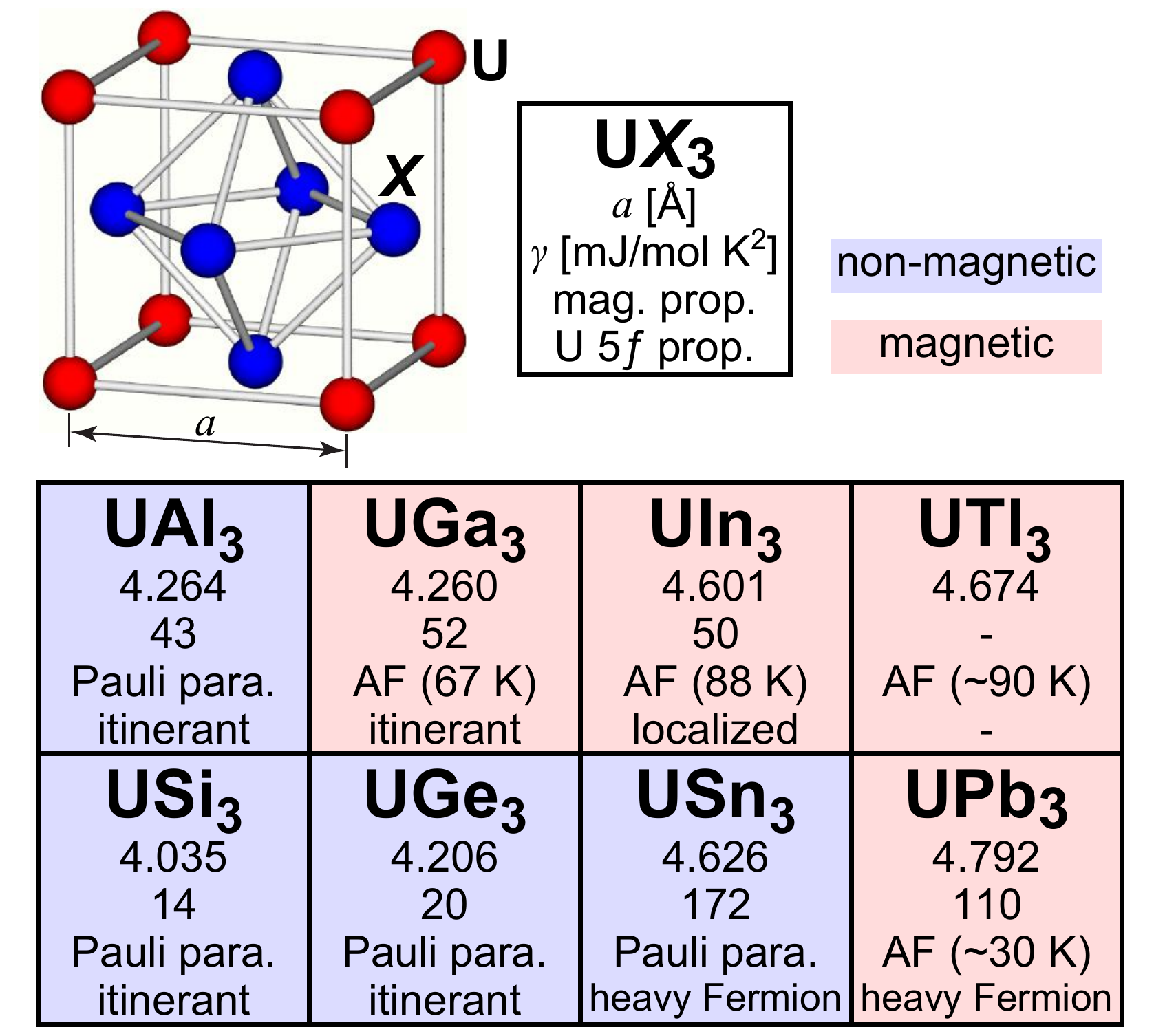}
	\caption{(Online color)
		Crystal structure and summary of physical properties of $\mathrm{U}X_3$ compounds \cite{UX3_SSC, UX3, UX3_dHvA,  Onuki_review_JPSJ, UIn3_NMR}.		
	}
\label{UX3_crystal}
\end{figure}

Magnetism in $f$-electron materials is a test stand of the modern concepts of magnetism.
Hybridization between $f$ electrons and ligand states results in a competition between itinerant and localized natures of $f$ electrons, which manifests as the complex magnetic behaviors of these compounds.
To unveil the microscopic origins of these magnetic properties, systematic control of $f$-ligand hybridization in $f$-electron materials is desirable. 
Binary uranium compounds with $\mathrm{U}X_3$ ($X$ is a group 13 or 14 element) stoichiometry and $\mathrm{AuCu_3}$-type crystal structure comprise an ideal model system to understand the relationship between hybridization and the origin of a rich variety of magnetic properties in uranium-based compounds.  
Figure~\ref{UX3_crystal} summarizes the physical parameters and properties of this series of compounds \cite{UX3_SSC, UX3, UX3_dHvA,  Onuki_review_JPSJ, UIn3_NMR}.
They exhibit various physical properties depending on $X$, and their specific heat coefficients range from $\gamma = 14~\mathrm{mJ/mol K^2}$ in $\mathrm{USi_3}$ to $\gamma =170~\mathrm{mJ/mol K^2}$ in $\mathrm{USn_3}$.
Their lattice constants are considerably larger than the Hill limit ($\sim3.4~\mathrm{\AA}$), suggesting that the hybridization between $f$-state and ligand $X$ states is a key parameter for these compounds.
Generally, as $X$ becomes heavier, the lattice constant increases, and the compound tends to be magnetic.
Among the considered series of compounds, \UAlGaIn has very different physical properties depending on $X$, namely, enhanced Pauli paramagnetism in \UAl \cite{UAl3_dHvA}, itinerant antiferromagnetism in \UGa \cite{UGa3_Kaczorowski}, and localized antiferromagnetism in \UIn \cite{UIn3_NMR}.
Therefore, they comprise an excellent model system to study the origin of magnetism in $f$-electron systems.

In the present work, we studied the electronic structures of \UAlGaIn to reveal the origin of their different magnetic properties by using valence-band and core-level photoelectron spectroscopies.
Furthermore, angle-resolved photoelectron spectroscopy (ARPES) with photon energies of \hn{=575-650} were performed for \UAl and \UGa.
Their detailed band structures and Fermi surfaces were revealed experimentally, and they were compared with the result of the band-structure calculation treating \Uf electrons as being itinerant.

\UAl is a paramagnetic compound with a relatively large specific heat coefficient of \gm{43} \cite{UX3_dHvA}.
In a dHvA study of \UAl, several branches were observed, and a few of them were explained by band-structure calculations based on itinerant \Uf states \cite{UX3_dHvA,UAl3_dHvA}.
In recent years, fully relativistic band-structure calculations could explain the origin of most of other branches, suggesting that \Uf electrons have very itinerant characteristics in this compound \cite{UAl3_Maehira}.

\UGa is an antiferromagnet with \TN{67} \cite{UGa3_AF}.
Its specific heat coefficient is slightly higher than that of \UAl.
Many experimental and theoretical studies on \UGa have suggested that \UGa is an itinerant antiferromagnet \cite{UGaSn3, UX3_dHvA, UGa3_optical, UGa3_Hiess, UGa3_Usuda, UGa3_positron}.
Its magnetic ordering is of type-II with the magnetic propagating vector $\vec{Q}=\lfloor \frac{1}{2} \frac{1}{2} \frac{1}{2} \rfloor$ \cite{UGa3_neutron}.
There exists another phase transition at $T=40~\mathrm{K}$, and it has been ascribed to the reorientation of magnetic moments, although the direction of magnetic moments has been controversial \cite{UGa3_NMR,UGa3_Mossbauer}.
Several experimental methods have been applied for studying electronic structure of \UGa such as resonant photoemission \cite{UGa3_RPES}, dHvA measurement \cite{UX3_dHvA,UGa3_Aoki}, positron annihilation \cite{UGa3_positron}, and magnetic X-ray scattering \cite{UGa3_MXS}.
Despite these extensive studies, the details of its electronic structure have not been well understood.
An interesting question is the difference in the magnetic properties of \UAl and \UGa.
They have very similar lattice constants, but \UAl is a paramagnet, while \UGa is an antiferromagnet.
Therefore, the magnetic ordering of \UGa originates from the tiny differences in their electronic structures.

\UIn is an antiferromagnet with \TN{88} \cite{UIn3_NMR}.
The lattice constants of \UAl and \UGa are almost identical, while that of \UIn is about 8 \% larger than theirs. 
This leads to the weaker hybridized nature of \Uf in \UIn compared to those in \UAl and \UGa.
$\mathrm{^{115}In}$-NMR and NQR studies revealed that \Uf electrons have localized natures well above the N\'eel temperature \cite{UIn3_NMR}.
They further suggested that a plausible ordering vector is along the $\lfloor 110 \rfloor$ direction.
Meanwhile, there are only a few studies on the electronic structure of \UIn.
Sarma \etal conducted a resonant photoemission study of $\mathrm{U(Sn,In)_3}$ and found that the spectral profiles of the \Uf contributions do not show any significant changes within the series \cite{UIn3_RPES}.
In a dHvA study of \UIn, several branches originating from closed Fermi surfaces and multiply-connected Fermi surfaces were observed \cite{UIn3_dHvA}.
The estimated electron masses of these branches were $10-33 m_0$, suggesting the existence of heavy quasi-particle bands.
By contrast, comparison with band-structure calculation has not been performed yet, and the overall topology of the Fermi surface is not well understood.
Therefore, knowledge about its electronic structure is very limited at present.

An interesting standpoint is that \UIn is considered as \UGa or \UAl under negative pressures.
Pressure is a clean tuning parameter for controlling the physical properties of strongly correlated materials.
In particular, it has been used to tune the electronic structure of $f$-electron materials to explore their quantum-criticality and unconventional superconductivities.
Meanwhile, the effect of pressure on their electronic structures has been not well studied because spectroscopic studies are very difficult to perform in high-pressure cells.
Therefore, \UAlGaIn can be used as a model system for studying the pressure effect by using photoelectron spectroscopy.

\section{EXPERIMENTAL PROCEDURES}
Photoemission experiments were performed at the soft X-ray beamline BL23SU in SPring-8 \cite{BL23SU,BL23SU2}.
High-quality single crystals were grown by the self-flux method, as described in Refs.~\cite{UAl3_dHvA,UGa3_Aoki,UIn3_dHvA}.
Clean sample surfaces were obtained by cleaving the samples {\it in situ} with the surface under an ultra-high vacuum (UHV) condition.
Among the series of compounds, we could not obtain a flat cleaving surface in the case of \UIn.
The ARPES spectra of \UIn were dominated by non-dispersive features that might have originated from elastically scattered photoelectrons from irregular surface.
Therefore, only angle-integrated photoemission spectra are shown in the present paper.
The overall energy resolution in the angle-integrated photoemission experiments at $h\nu = 800$ and $850~\mathrm{eV}$ was about $120~\mathrm{meV}$, while that in the ARPES experiments at \hn{=575-650} was about $100~\mathrm{meV}$.
The position of \EF was determined carefully by measuring of the vapor-deposited gold film.
During the measurements, the vacuum was typically $<1 \times 10^{-8}~\mathrm{Pa}$, and the sample surfaces were stable for the duration of measurements ($1-2$ days) because no significant changes were observed in the ARPES spectra during the measurement period.
The positions of ARPES cuts were determined by assuming a free-electron final state with an inner potential of $V_{0}=12~\mathrm{eV}$.
Background contributions in ARPES spectra originated from elastically scattered photoelectrons due to surface disorder or phonons were subtracted by assuming momentum-independent spectra.
The details of the procedure are described in Ref.~\cite{UGe2_UCoGe_ARPES}. 

\section{RESULTS}
\begin{figure}
	\includegraphics[scale=0.5]{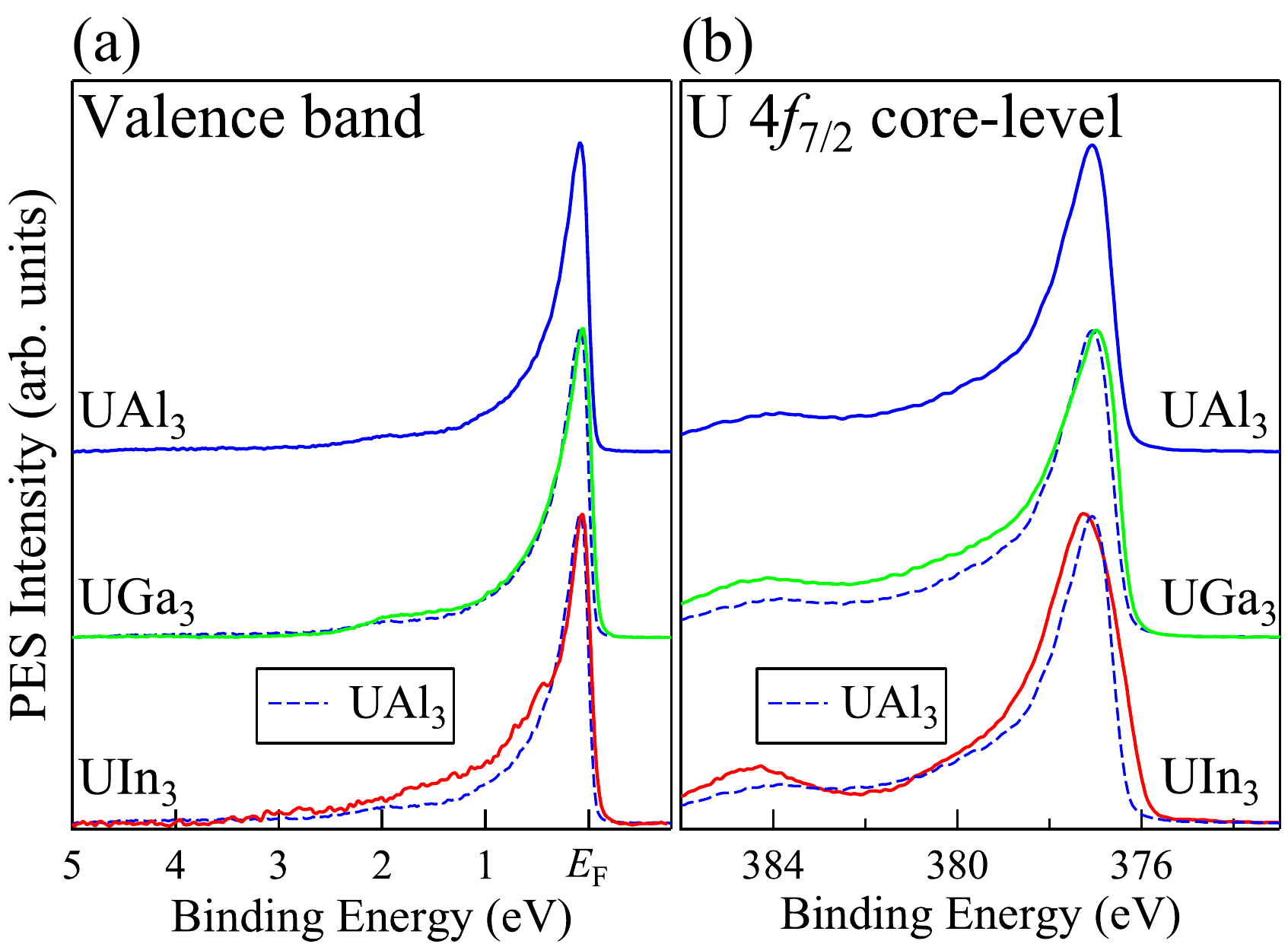}
	\caption{(Online color)
		Angle-integrated photoemission spectra of \UAlGaIn.
		(a) Valence band spectrum of \UAlGaIn measured at \hn{=800}.
		The spectrum of \UAl is superimposed onto the spectra of \UGa and \UAl.
		(b) \orb{U}{4f} core-level spectra of \UAlGaIn.
		The spectra of \UAl and \UGa were measured at \hn{=800} while that of \UIn was measured at \hn{=850} to avoid the contribution of $\mathrm{In}$-originated Auger signals, which overlap with \orb{U}{4f} core-level.
		The spectrum of \UAl was superimposed onto the spectra of \UGa and \UIn as shown by the dotted line.
	}
\label{AIPES}
\end{figure}

\subsection{Angle-integrated photoemission spectra of \UAlGaIn}
Figure~\ref{AIPES} summarizes the angle-integrated photoemission spectra of \UAlGaIn.
Figure~\ref{AIPES}(a) shows the valence band spectra of \UAlGaIn.
The spectra were measured at $20~\mathrm{K}$, and \UAl was in the paramagnetic phase while \UGa and \UIn were in the antiferromagnetic phase.
According to the calculated cross-sections of atomic orbitals \cite{Atomic}, the cross-section of \Uf orbitals is more than one-order larger than those of \orb{Al}{3s, p}, \orb{Ga}{4s, p}, and \orb{In}{5s, p} orbitals.
Therefore, the signals from \Uf states are dominant in these spectra.
These spectra exhibit an asymmetric shape with a sharp peak just below the Fermi energy.
Their spectral profiles are very similar to those of itinerant uranium compounds, such as $\mathrm{UB_2}$ \cite{UB2_ARPES} and $\mathrm{UN}$ \cite{UN_ARPES}, suggesting that \Uf states have an itinerant character in these compounds.
By the contrast, the spectral profile of \UIn is slightly different from those of \UAl and \UGa.
To understand the differences in their spectral profiles, we superimposed the spectrum of \UAl on the spectra of \UGa and \UIn.
The spectrum of \UGa is almost identical to that of \UAl, while that of \UIn has a shoulder structure at \EB{\sim 0.5} whose tail extends \EB{\sim 2}.
Note that the structure cannot be due to the antiferromagnetic transition because a similar structure does not exist in the valence band spectrum of \UGa, which was also measured in the antiferromagnetic phase.
Therefore, this shoulder structure has originated from the incoherent part of the correlated \Uf states.
A similar satellite structure was observed in the valence band spectrum of $\mathrm{UBe_{13}}$ \cite{Laub_UBe13} which is also considered as the contribution of correlated \Uf states.

\begin{figure*}
	\includegraphics[scale=0.5]{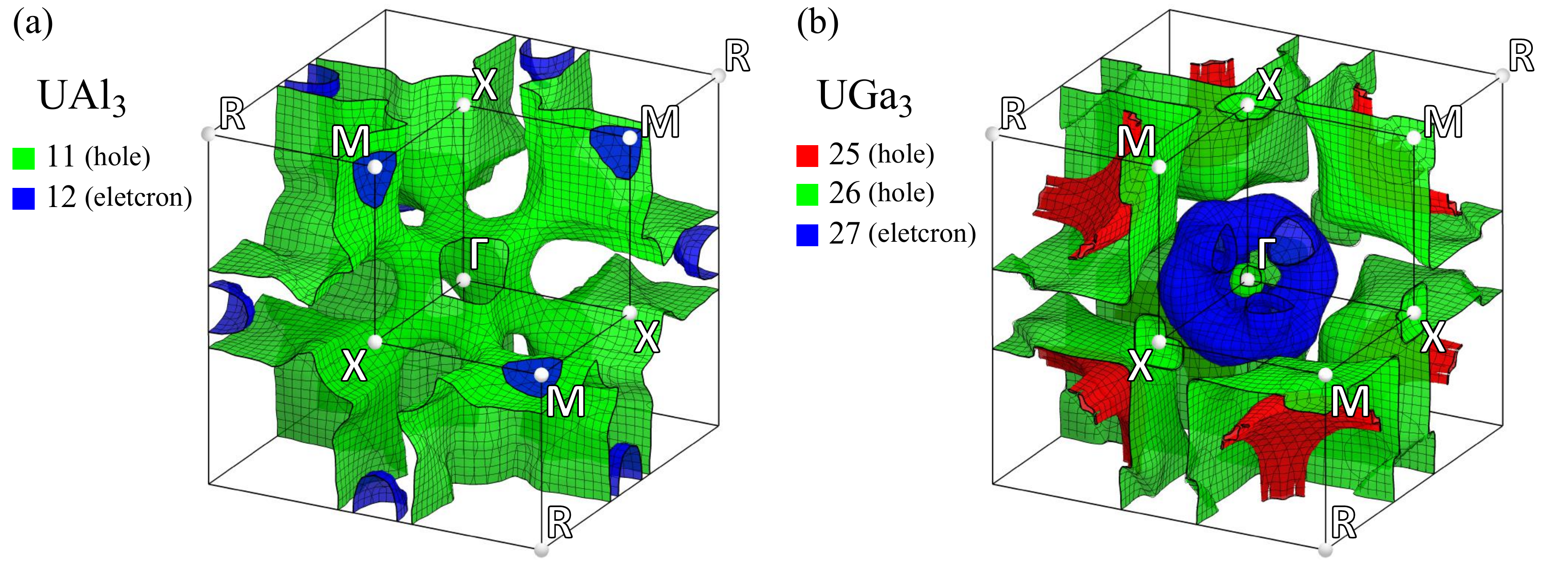}
	\caption{(Online color)
		Three-dimensional shape of calculated Fermi surface of \UAl and \UGa.
		(a) Calculated Fermi surface of \UAl.
		Two kinds of Fermi surfaces are expected in \UAl.
		Band 11 forms a multiply-connected Fermi surface consisting of a large Fermi surface centered at the \Gm and the \pnt{R}points.
		Band 12 forms a small electron-type Fermi surface centered at the \pnt{M} point.
		(b) Calculated Fermi surface of \UGa.
		Three bands are expected to form Fermi surfaces in \UGa.
		Band 25 forms a thin closed Fermi surface around the \pnt{R} point.
		Band 26 forms a large hole-type Fermi surface centered at the \pnt{R} point and two small Fermi surfaces centered at the \Gm and \pnt{X} points.
		Band 27 gives rise to a hollow centered spherical Fermi surface centered at the \Gm point.
	}
\label{3DFS}
\end{figure*}
To further understand the nature of \Uf states in these compounds, we measured the $\mathrm{U}~4f_{7/2}$ core-level spectra of \UAlGaIn, which are shown in Fig.~\ref{AIPES}(b).
These spectra were replotted from Ref.~\cite{SF_review_JPSJ}. 
The core-level spectrum is a sensitive probe of the local electronic structure \cite{Ucore, SF_review_JPSJ}.
The spectra of these compounds have an asymmetric line shape with a tail toward the higher binding energies.
In addition to the main line located at \EB{= 377.0-377.3}, a weak satellite structure can be observed in the high-binding-energy side of the main line at about $7~\mathrm{eV}$ (\EB{\sim 384}).
This is called as the $7~\mathrm{eV}$ satellite~\cite{Laub_U4f}, and it has been observed in many uranium based compounds \cite{Ucore, SF_review_JPSJ}.
The core-level spectrum of \UAl was superimposed on the spectra of \UGa and \UIn, as in the case of their valence band spectra. 
The spectra exhibit considerable differences.
The tail of the main line of \UGa is more enhanced than that of \UAl, suggesting that the asymmetry of the main line is larger than the case of \UAl.
In general, the asymmetry of the main line is associated with the density of states (DOS) at \EF in itinerant uranium compounds \cite{Ucore}.
Compounds with higher DOSs at \EF have main lines with greater asymmetry.
Therefore, the larger asymmetry in the main line of \UGa than that of \UAl suggests that \UGa has higher DOS at \EF than that does \UAl.
By contrast, the spectrum of \UIn is broadened and is located on the higher-binding-energy side.
In addition, its spectral shape becomes more symmetric.
A detailed analysis of the main line of \UIn suggests that it consists of two components, and the one on the high-binding-energy side becomes dominant \cite{SF_review_JPSJ}.
This is a characteristic feature of the main lines of localized compounds, suggesting that \Uf in \UIn is more localized than those in \UAl and \UGa.
The intensities of the satellite structures of \UAl and \UGa are similar to those of other itinerant compounds \cite{Ucore, SF_review_JPSJ}, but that of \UIn is more enhanced than those of \UAl and \UGa.
In general, the intensity of the satellite structure is more enhanced in localized compounds \cite{Ucore}, and this slightly enhanced satellite intensity also indicates that \Uf electrons are more localized in \UIn than in \UAl and \UGa.
It should be noted that there are some theoretical attempts to reproduce these structures by the single impurity Anderson model \cite{Okada_core,Zwicknagl_core}.
A systematic trend observed in the valence band and core-level spectra of these compounds would be helpful to further understand the microscopic origin of these satellite structures.

Accordingly, both valence band and core-level spectra indicate that the \Uf states of \UAl and \UGa have an essentially itinerant character.
\UAl and \UGa have similar degrees of delocalization of the \Uf states, although the correlation effect is somewhat enhanced in \UGa.
By contrast, the nature of the electronic structure of \UIn is clearly different.
The valence band spectrum of \UIn is accompanied with the satellite structure, and the main lines and satellite structures in the core-level spectra are also different from those of \UAl and \UGa.
This should be due to the reduced hybridization in \UIn originated from its larger lattice constant than those of \UAl and \UGa.

\subsection{Band-structure calculation}
Before showing the experimental ARPES spectra, we overview the result of our band-structure calculations of \UAl and \UGa.
Figure~\ref{3DFS} shows the calculated Fermi surfaces of \UAl and \UGa in the paramagnetic phase.
In the present study, band-structure calculation in the paramagnetic phase is used for comparison between ARPES spectra and the calculation of \UGa because the changes in the spectral profiles due to antiferromagnetic transition are very small, as discussed in Sec.~\ref{UGa3_AF}.

Figure~\ref{3DFS} (a) shows the calculated Fermi surfaces of \UAl.
The calculated Fermi surface of \UAl consists of multiply-connected hole-type Fermi surfaces formed by band 11 and small electron pocket formed by band 12.
The Fermi surface formed by band 11 consists of two large Fermi surfaces centered at the \Gm and the \pnt{R} points, and they are connected along the $\lfloor 111 \rfloor$ direction.
Note that the topology of the calculated Fermi surface is basically consistent with that obtained from previous band-structure calculations \cite{UX3_dHvA,UAl3_dHvA,UAl3_Maehira}, although there exist minor differences.
For example, the Fermi surfaces centered at \Gm and \pnt{R} points are not connected in Ref.~\cite{UAl3_dHvA}.
Furthermore, the electron pocket centered at the \pnt{M} point does not exist in these previous calculations.
These differences might have originated from tiny differences in the band structure near \EF.
There exist very narrow bands with energy dispersions of less than 50 meV in these calculations, and tiny changes in the structures of these bands due to different computational factors can alter the shapes of Fermi surface very easily.
Experimentally, several branches originating from this Fermi surface were observed by dHvA measurement \cite{UAl3_dHvA}, and they were interpreted reasonably as signals from the two large Fermi surfaces centered at the \Gm and the \pnt{R} points.

\begin{figure*}[ht]
	\includegraphics[scale=0.5]{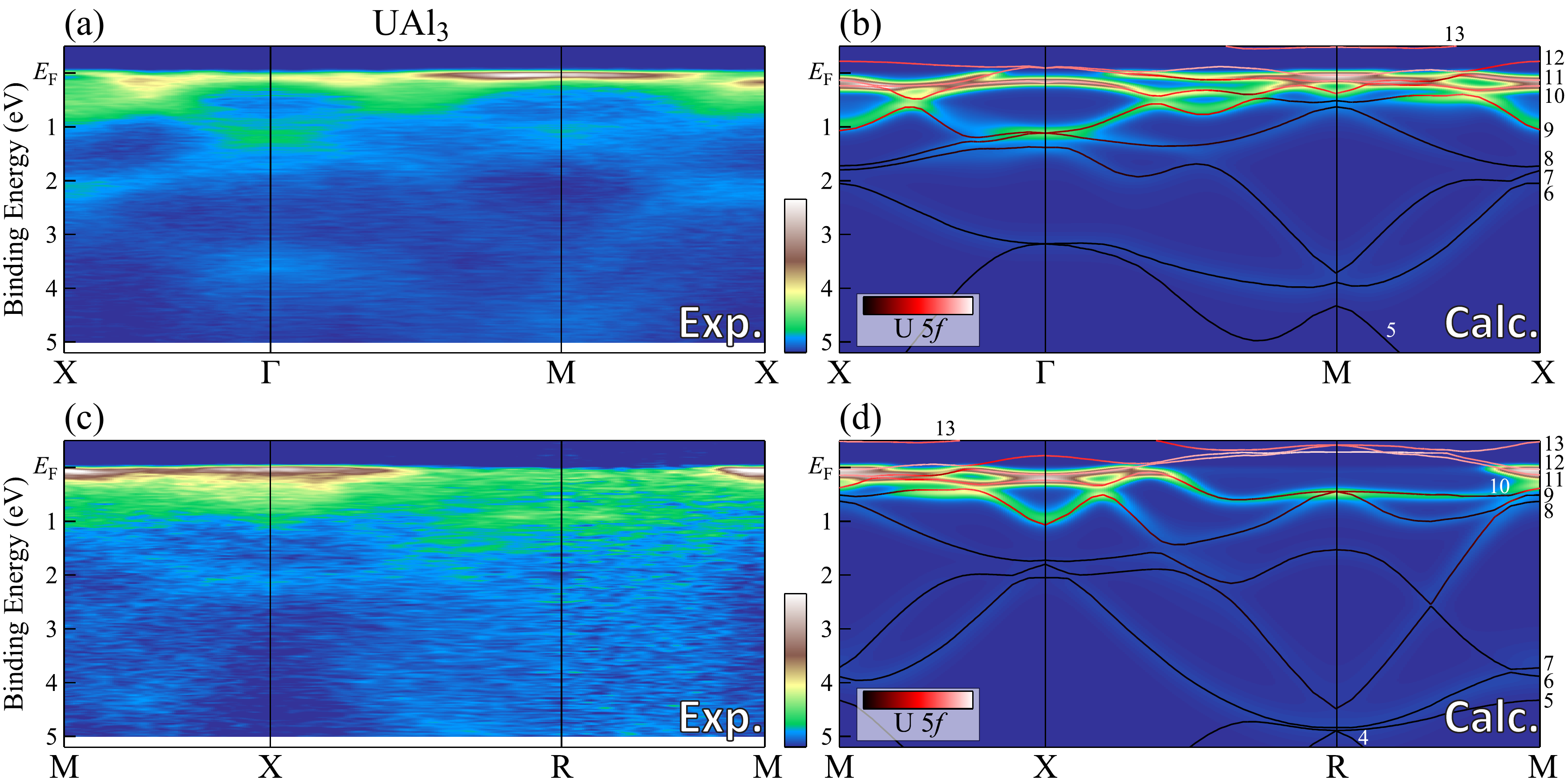}
	\caption{(Online color)
		Experimental band structure of \UAl and result of band-structure calculation.
		(a) Band structure of \UAl along the $\mathrm{X}-\mathrm{\Gamma}-\mathrm{M}-\mathrm{X}$ high-symmetry line obtained from ARPES spectra measured at \hn{=645}.
		(b) Calculated band structure and simulated ARPES spectra based on band-structure calculation along the $\mathrm{X}-\mathrm{\Gamma}-\mathrm{M}-\mathrm{X}$ high-symmetry line.
		(c) Band structure of \UAl along the $\mathrm{M}-\mathrm{X}-\mathrm{R}-\mathrm{M}$ high-symmetry line obtained from ARPES spectra measured at \hn{=575}.
		(d) Calculated band structure and simulated ARPES spectra based on band-structure calculation along the $\mathrm{M}-\mathrm{X}-\mathrm{R}-\mathrm{M}$ high-symmetry line.
	}
\label{UAl3_band}
\end{figure*}
Figure~\ref{3DFS} (b) shows the calculated Fermi surfaces of \UGa.
The topologies of the calculated Fermi surfaces of \UGa are very similar to those of \UAl, but those of \UGa have considerably complicated structures.
Band 25 forms a thin cubic frame-like Fermi surface centered at the \pnt{R} point which has no equivalent in the Fermi surface of \UAl.
Band 26 forms a cubic Fermi surface centered at the \pnt{R} point, which is very similar to the Fermi surface formed by band 11 in the case of \UAl, although its shape is closer to cubic in the case of \UGa.
It also forms a small hole pocket at the \Gm point, but the size of which is considerably smaller than the corresponding hole pocket in \UAl.
As a result, the large Fermi surface centered at the \pnt{R} and the hole pocket centered at the \Gm point are disconnected in \UGa.
Band 27 forms a hollow spherical Fermi surface around the \Gm point, which has no equivalent in the Fermi surface in \UAl.
The topology of these calculated Fermi surfaces is essentially identical to the topology obtained in previous calculations \cite{UGa3_positron}, although there are some minor differences.
dHvA measurement of \UGa was performed, and the branches observed were compared with the results of band-structure calculation in the antiferromagnetic phase, but any satisfactory agreement was not obtained between them \cite{UGa3_Aoki}.
A positron annihilation study suggested overall agreement between the experimental data and the band-structure calculation in the paramagnetic phase, but the details of the Fermi surface were not understood \cite{UGa3_positron}.
Therefore, information about the Fermi surface of \UGa is very limited.

\subsection{Band structure and Fermi surface of \UAl}
We begin with the overall band structure of \UAl.
In Fig.~\ref{UAl3_band}, we summarize the experimental ARPES spectra and the result of the band-structure calculation of \UAl along several high-symmetry lines.
Figure~\ref{UAl3_band} (a) shows the ARPES spectra of \UAl measured along the $\mathrm{X} - \mathrm{\Gamma}-\mathrm{M} - \mathrm{X}$ high-symmetry line at the photon energy of \hn{=645}.
Note that the locations of the \pnt{X} point in the leftmost and the rightmost sides have different measurement configurations, and the spectra show different profiles owing to matrix element effects.
Clear energy dispersions can be observed.
The strongly dispersive features corresponding to the higher-binding energy of \EB{>1} are mainly the contributions of the $\mathrm{Al}~3s, 3p$ states.
By contrast, the less dispersive features near \EF are quasi-particle bands with dominant contribution of the \Uf states.
They have finite energy dispersions, and form the Fermi surface of \UAl.
Figure~\ref{UAl3_band} (b) shows the band dispersions and the simulated ARPES spectra based on the band-structure calculation along the same high-symmetry lines.
The color coding of the bands is the projection of the contributions of the \Uf states.
The overall experimental band structure is well explained by the band-structure calculation.
The dispersive features located in the higher binding energies correspond well with bands 5--8 in the calculation.
The feature near \EF seems to correspond to bands 9--12.
The overall shapes of these bands agree reasonably well between experimental and calculation results.
Figure~\ref{UAl3_band} (c) shows the experimental ARPES spectra of \UAl measured along the $\mathrm{M}-\mathrm{X}-\mathrm{R}-\mathrm{M}$ high-symmetry line at the photon energy of \hn{=575}.
There exist similar types of energy dispersions to the spectra shown in Fig.~\ref{UAl3_band} (a), and their overall structure can be explained by the band-structure calculation and the simulated ARPES spectra shown in Fig.~\ref{UAl3_band}(d).

\begin{figure*}[ht]
	\includegraphics[scale=0.5]{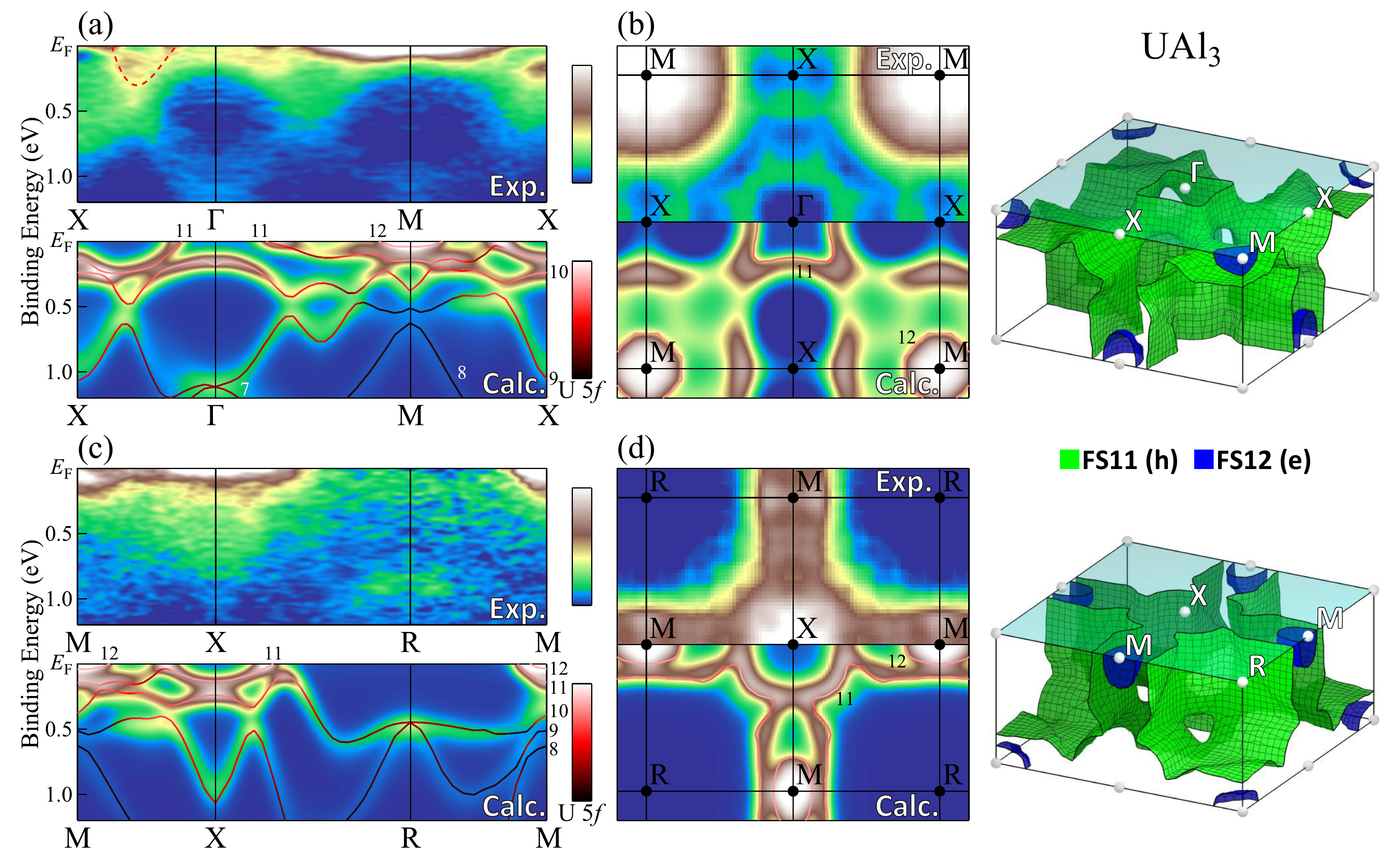}
	\caption{(Online color)
		Band structure and Fermi surface of \UAl.
		(a) Comparison of ARPES spectra measured along $\mathrm{X} - \mathrm{\Gamma} - \mathrm{M} - \mathrm{X}$ high-symmetry line and corresponding result of band-structure calculation near \EF.
		The spectra are divided by the Fermi--Dirac function to clearly show the structure just below \EF.
		Dashed line is the guide to eye. 
		(b) Comparison of experimental Fermi surface map (upper half) and calculated Fermi surface (lower half) within the $\mathrm{\Gamma} - \mathrm{M} - \mathrm{X}$ plane, and three-dimensional Fermi surface.
		(c) Same as (a) but along the $\mathrm{M} - \mathrm{X} - \mathrm{R} - \mathrm{M}$ high-symmetry line.
		(d) Same as (b) but within the $\mathrm{X} - \mathrm{R} - \mathrm{M}$ plane.
	}
\label{UAl3_band_FS}
\end{figure*}
To further understand the electronic structure near \EF, blow-up of the ARPES spectra and the experimental and calculated Fermi surfaces of \UAl are shown in Fig.~\ref{UAl3_band_FS}.
Figure~\ref{UAl3_band_FS} (a) shows a comparison of the ARPES spectra and the calculated energy dispersions together with the simulated ARPES spectra along the $\mathrm{X} - \mathrm{\Gamma} - \mathrm{M} - \mathrm{X}$ high-symmetry line.
These spectra are divided by the Fermi -- Dirac function broadened by the instrumental energy resolution to avoid the influences of Fermi cut-off.
There exist one-to-one correspondences between the experimentally observed bands and the calculated bands.
Band 11 forms a hole pocket centered at the \Gm point in the calculation, and there exist very similar structures in the experimental spectra.
In addition, a hole pocket centered at the \pnt{X} point is recognized in the profiles at the leftmost \pnt{X} point of the experimental spectra, but there is no corresponding Fermi surface in the calculated results.
Although the experimental spectra around the \pnt{M} point are more featureless than the calculated result, there exists a similar high-intensity region near \EF in the calculation.
Figure~\ref{UAl3_band_FS} (b) shows a comparison between the experimental Fermi surface map (upper panel) and the simulated ARPES spectra (lower panel).
Both the experimental and the calculated Fermi surface maps have very similar features.
Especially, very similar squared features centered at the \Gm point were observed in both the experimental and the calculated Fermi surface map.
By contrast, the experimental and calculated features around the \pnt{M} point are somewhat different.
The experimental Fermi surface map has a large circular region with enhanced intensity, while the calculation predicts a more complicated structure consisting of a small circular region with enhanced intensity at the \pnt{M} point surrounded by arcs.
This difference originates from plainer feature of bands near \EF around the \pnt{M} point in the experiment than in the calculation as shown in Fig.~\ref{UAl3_band_FS} (a).
This might be due to the renormalization of experimental band corresponding to band 11, which leads to the featureless structure in the experimental Fermi surface map.

\begin{figure*}[ht]
	\includegraphics[scale=0.5]{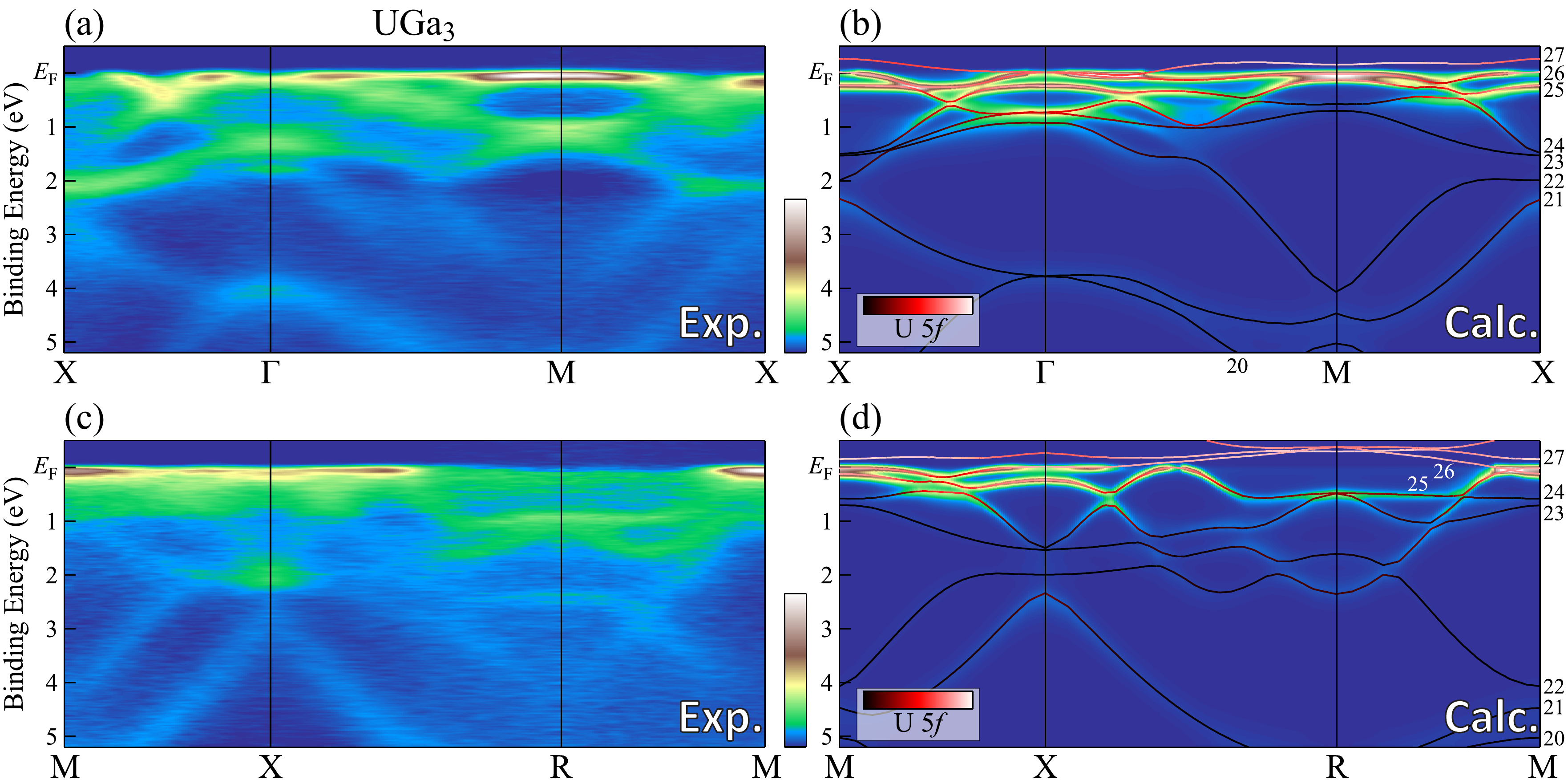}
	\caption{(Online color)
		Experimental band structure of \UGa, and result of band-structure calculation.
		(a) Band structure of \UGa along the $\mathrm{X}-\mathrm{\Gamma}-\mathrm{M}-\mathrm{X}$ high-symmetry line obtained from ARPES spectra measured at \hn{=650}.
		(b) Calculated band structure and simulated ARPES spectra based on band-structure calculation along the $\mathrm{X}-\mathrm{\Gamma}-\mathrm{M}-\mathrm{X}$ high-symmetry line.
		(c) Band structure of \UGa along the $\mathrm{M}-\mathrm{X}-\mathrm{R}-\mathrm{M}$ high-symmetry line obtained by ARPES spectra measured at \hn{=580}.
		(d) Calculated band structure and the simulated ARPES spectra based on band-structure calculation along the $\mathrm{M}-\mathrm{X}-\mathrm{R}-\mathrm{M}$ high-symmetry line.
	}
	\label{UGa3_band}
\end{figure*}
Figure~\ref{UAl3_band_FS} (c) shows the same comparison of Fig.~\ref{UAl3_band_FS} (a), but along the $\mathrm{M} - \mathrm{X} - \mathrm{R} - \mathrm{M}$ high-symmetry line.
Very similar correspondence between the experimental and the calculated results can be seen in these spectra.
There is overall agreement between the experimental and the calculated results.
Especially, the feature around the \pnt{M} point shows good agreement among them.
In the calculation, two bands form Fermi surfaces around the \pnt{M} point along the $\mathrm{R} - \mathrm{M}$ high-symmetry line, but they cannot be resolved in the experimental spectra.
Meanwhile, as in the case of the $\mathrm{X} - \mathrm{\Gamma} - \mathrm{M} - \mathrm{X}$ high-symmetry line shown in Fig.~\ref{UAl3_band_FS} (a), bands near \EF are pushed toward the lower-binding-energy side, and their profiles become more featureless.
Figure~\ref{UAl3_band_FS} (d) shows a comparison between the experimental Fermi surface map and the simulated ARPES spectra.
There is a similar relationship between the experimental Fermi surface map and the result of the band-structure calculation within the $\mathrm{\Gamma} - \mathrm{M} - \mathrm{X}$ high-symmetry plane shown in Fig.~\ref{UAl3_band_FS} (b).
Although the overall shape of the features is very similar between the experiment and the calculation, the calculated map has a  more complicated structure.
This more featureless nature of the experimental Fermi surface map might be also due to the renormalized nature of bands near \EF.
Nevertheless, the feature corresponding to the cubic Fermi surface centered at the \pnt{R} point is observed experimentally, which corresponds well to the Fermi surface formed by band 11 in the calculation.

Accordingly, the band structure and the Fermi surface of \UAl were explained well by the band-structure calculation although the bands near \EF were renormalized considerably.
Especially, the topology of the Fermi surface is mostly identical to the result of the calculation, although there are a few minor differences.

\begin{figure*}[ht]
	\includegraphics[scale=0.5]{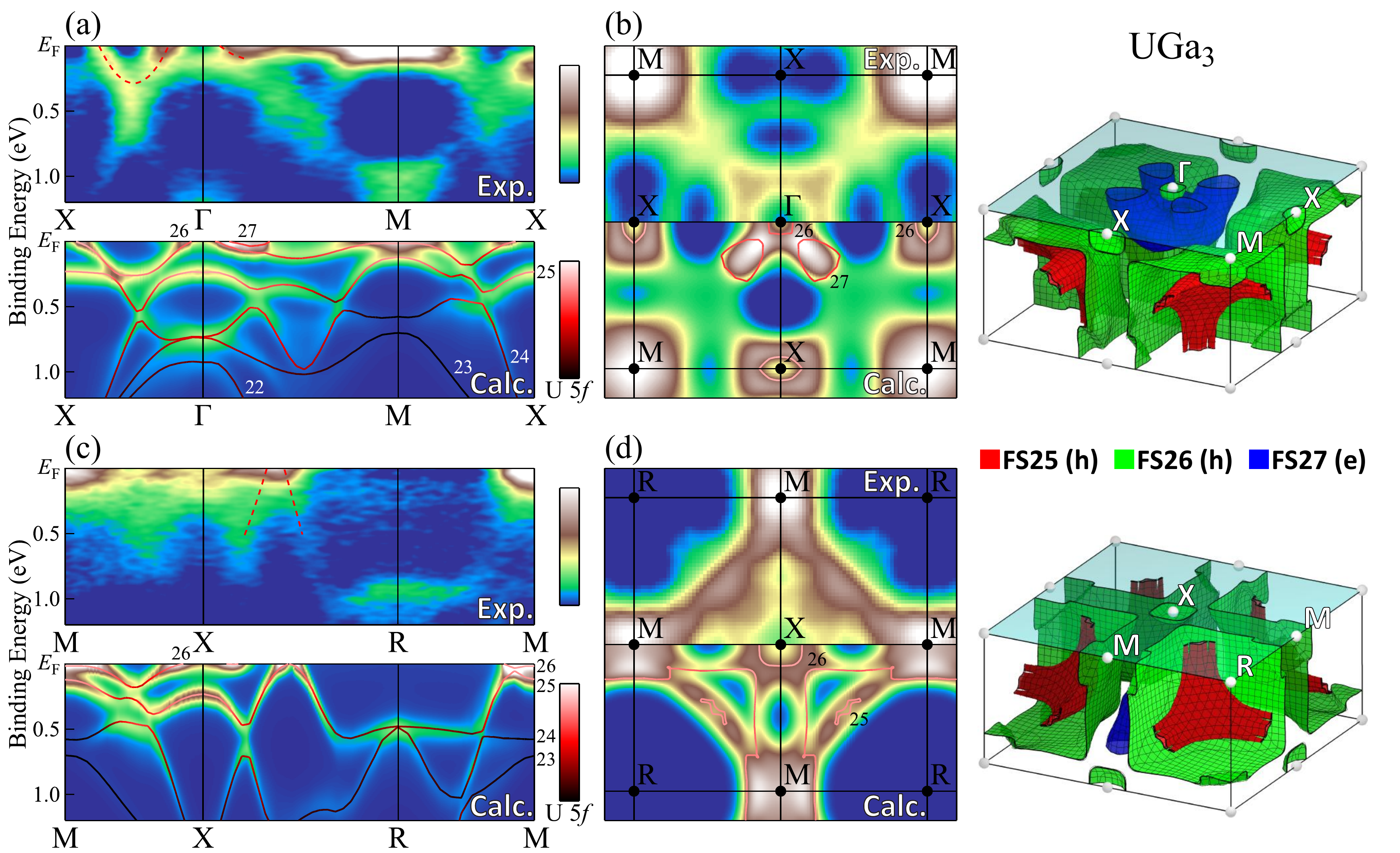}
	\caption{(Online color)
		Band structure and Fermi surface of \UGa.
		(a) Comparison of ARPES spectra measured along the $\mathrm{X} - \mathrm{\Gamma} - \mathrm{M} - \mathrm{X}$ high-symmetry line and corresponding result of band-structure calculation near \EF.
		The spectra are divided by the Fermi--Dirac function to clearly show the structure just below \EF.
		Dashed line is the guide to eye. 
		(b) Comparison of the experimental Fermi surface map (upper half) and calculated Fermi surface (lower half) within the $\mathrm{\Gamma} - \mathrm{M} - \mathrm{X}$ plane, and three-dimensional Fermi surface.
		(c) Same as (a) but along the $\mathrm{M} - \mathrm{X} - \mathrm{R} - \mathrm{M}$ high-symmetry line.
		(d) Same as (b) but within the $\mathrm{X} - \mathrm{R} - \mathrm{M}$ plane.
	}
	\label{UGa3_band_FS}
\end{figure*}
\subsection{Band structure and Fermi surface of \UGa}
We summarize the the experimental ARPES spectra of \UGa measured along several high-symmetry lines in Fig.~\ref{UGa3_band}.
Figure~\ref{UGa3_band} (a) shows the experimental ARPES spectra along the $\mathrm{X} - \mathrm{\Gamma} - \mathrm{M} - \mathrm{X}$ high-symmetry line measured at \hn{=650}.
The spectra exhibit clear energy dispersions, and their overall structure is very similar to the structure in the spectra of \UAl shown in Fig.~\ref{UAl3_band}.
There are dispersive bands with weak intensity on the high binding energy side, and they are ascribed to the \orb{Ga}{4s, p} states.
Less dispersive bands with enhanced intensity located near \EF are ascribed to \Uf states, which form narrow quasi-particle bands.
Figure~\ref{UGa3_band} (b) shows the band structure and the simulated ARPES spectra based on the band-structure calculation.
The experimental ARPES spectra are explained quantitatively by the calculated results.
Bands 20--24 consist mainly of the \orb{Ga}{4s,p} states, and they have one-to-one correspondence with the experimentally observed band dispersions.
Figure~\ref{UGa3_band} (c) shows the experimental ARPES spectra of \UGa along the $\mathrm{M}-\mathrm{X}-\mathrm{R}-\mathrm{M}$ high-symmetry line measured at \hn{=580}.
The nature of energy dispersion is very similar to that in the case of the $\mathrm{X}-\mathrm{\Gamma}-\mathrm{M}$ high-symmetry line shown in Fig.~\ref{UGa3_band} (a).
Figure~\ref{UGa3_band} (d) shows the calculated band structure and the simulated ARPES spectra based on the band-structure calculation.
There is overall agreement between the experimental and the calculated results as in the case of the $\mathrm{X} - \mathrm{\Gamma} - \mathrm{M} - \mathrm{X}$ high-symmetry shown in Figs.~\ref{UGa3_band} (a) and (b).
In particular, bands 20--24 correspond well with the experimental ARPES spectra shown in Fig.~\ref{UGa3_band} (c).
Note that overall band structures of \UAl and \UGa are very similar to each other.
Especially, the dispersive bands of \UAl and \UGa at high binding energies (\EB{\gtrsim 1}) show apparent one-to-one correspondences.

\begin{figure*}
	\includegraphics[scale=0.5]{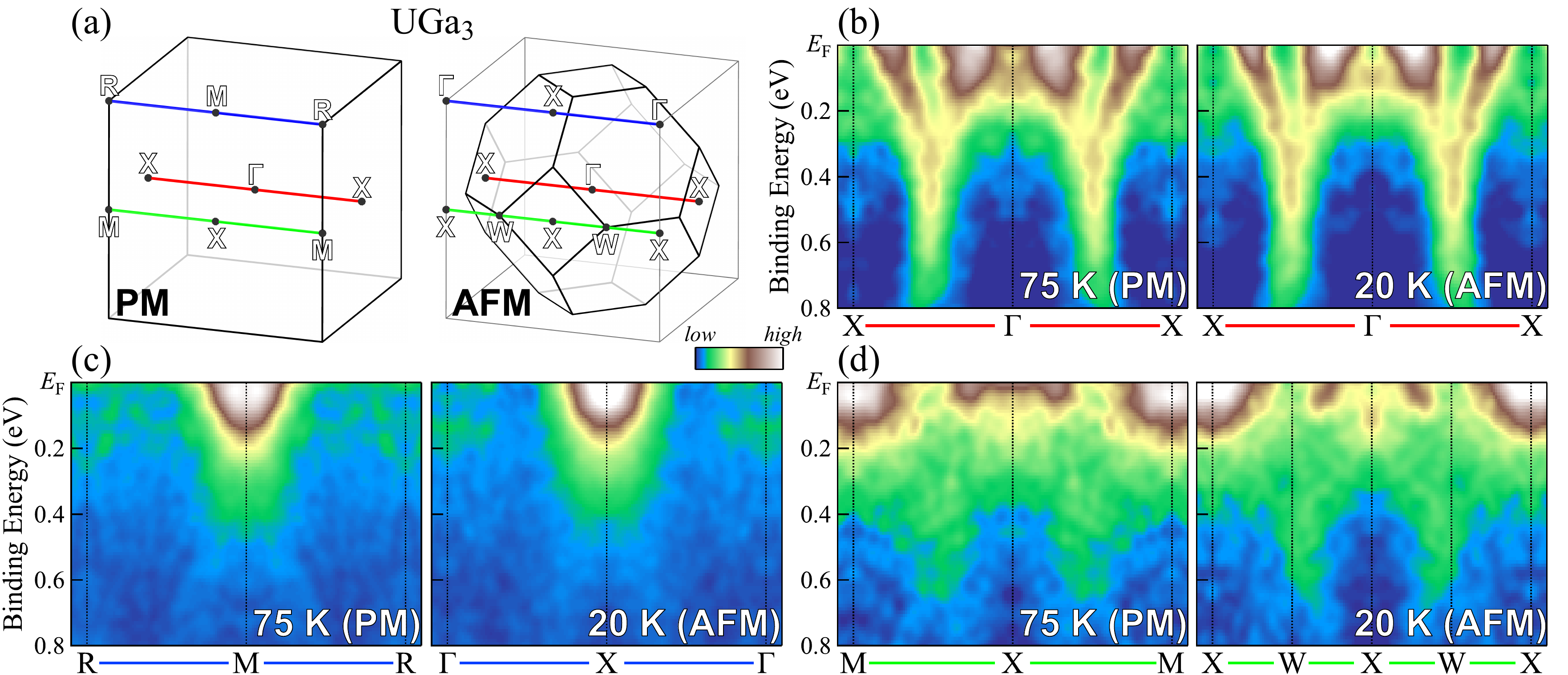}
	\caption{(Online color)
		Comparison of ARPES spectra of \UGa measured in the paramagnetic  and antiferromagnetic phases.
		(a) Brillouin zones of \UGa in the paramagnetic and antiferromagnetic phases.
		The symmetry of the Brillouin zone is simple cubic in the paramagnetic phase while it is face-centered cubic in the antiferromagnetic phase.
		(b) Comparison of ARPES spectra of \UGa measured in the paramagnetic phase ($75~\mathrm{K}$) and antiferromagnetic phase ($20~\mathrm{K}$).
		The scan corresponds to the \lineXXX{X}{\Gamma}{X} high-symmetry line in both the paramagnetic phase and the antiferromagnetic phase.
		(c) Same as (b), but the scan corresponds to the \lineXXX{R}{M}{R} high-symmetry line in the paramagnetic phase and \lineXXX{\Gamma}{X}{\Gamma} high-symmetry line in the antiferromagnetic phase.
		(d) Same as (b), but the scan corresponds to the \lineXXX{M}{X}{M} high-symmetry line in the paramagnetic phase and the $\mathrm{X} - \mathrm{W}- \mathrm{X} - \mathrm{W}- \mathrm{X}$ high-symmetry line in the antiferromagnetic phase.
	}
	\label{UGa3_tmpdep}
\end{figure*}
We further focus on the electronic structure near \EF.
Figure~\ref{UGa3_band_FS} summarizes the band structure near \EF and the Fermi surface of \UGa.
Figure~\ref{UGa3_band_FS} (a) shows a comparison between the experimental ARPES spectra and the calculated energy dispersions together with simulated ARPES spectra along the $\mathrm{X} - \mathrm{\Gamma} - \mathrm{M} - \mathrm{X}$ high-symmetry line.
The behavior of the quasi-particle bands near \EF can be recognized well from this comparison.
A parabolic dispersion forming the Fermi surface is observed clearly in the middle of the \lineXX{X}{\Gamma} high-symmetry line, and this band corresponds well to the calculated band 26. 
On the other hand, along the \lineXX{\Gamma}{M} high-symmetry line, there is hole-type energy dispersion around the \Gm point.
The band-structure calculation shows hole-type dispersions formed by band 26 and small electron pockets formed by band 27 around the area.
The experimentally observed dispersion corresponds well to the calculated band 26.
Because band 27 forms very small electron pocket, it is not clear whether the corresponding band exists in the experimental ARPES spectra.
The structure around the \pnt{M} point is very similar between experimental and calculated results although its detail was not well resolved in the experimental ARPES spectra.
Note that the intensity at \EF at the middle of the \lineXX{\Gamma}{M} high-symmetry line is featureless while the calculation predicts an energy dispersion of about $\sim 0.1~\mathrm{eV}$.
This difference can be understood that the experimental feature corresponds to the calculated band 26, but it is pushed toward \EF.
Therefore, the band forms Fermi surface similar to the calculated band 26, but it is strongly renormalized owing to the electron correlation effect.
To summarize, the experimental Fermi surface map is reasonably well explained by the band-structure calculation although the bands near \EF are renormalized.

Figure~\ref{UGa3_band_FS} (b) shows a comparison between the experimental Fermi surface maps obtained by integrating $100~\mathrm{meV}$ over \EF of ARPES spectra, and the result of the band-structure calculation within the $\mathrm{\Gamma}-\mathrm{M}-\mathrm{X}$ plane.
There is reasonable agreement between experimental map and the simulation results.
Meanwhile, the size of the hole pocket formed by band 26 around the \pnt{X} point is smaller than the one deduced from the experimental Fermi surface map.
The size of the hole pocket around the \Gm point is also smaller in the calculation than that in the experiment.
The existence of the small Fermi surface formed by band 27, which is located at the middle of $\mathrm{\Gamma-X}$ high-symmetry line, is not clear in the experimental Fermi surface map, but there exists a similar high intensity region along the \lineXX{\Gamma}{M} high-symmetry line, suggesting there might be a similar Fermi surface in the experiment.

Figure~\ref{UGa3_band_FS} (c) is identical to Fig.~\ref{UGa3_band_FS} (a) but along the $\mathrm{M}-\mathrm{X}-\mathrm{R}-\mathrm{M}$ high-symmetry line.
The experimental ARPES spectra are very similar to those of \UAl shown in Fig.~\ref{UAl3_band_FS} (c).
There exists a hole-like dispersion along the \lineXX{X}{R} high-symmetry line as indicated by dotted curves. 
Furthermore, the intensity at \EF is particularly enhanced around the \pnt{M} point, implying that there exist Fermi surfaces.
In addition, the intensity at \EF is enhanced around the \pnt{M} point, suggesting that there exists some Fermi surfaces around the \pnt{M} point as in the case of \UAl.
In the calculation, bands 25 and 26 form Fermi surfaces.
Band 25 forms a hole pocket in the middle of the \lineXX{X}{R} high-symmetry line, while band 26 forms a tiny hole pocket around the \pnt{X} point and large cubic Fermi surface around the \pnt{R} point as shown in the three-dimensional Fermi surface.
The hole pocket formed by band 25 corresponds well to the experimentally observed hole pocket along the \lineXX{X}{R} high-symmetry line.
In addition, the feature formed by band 26 also has a good correspondence to the experimentally observed feature in the vicinity of \EF although their details were not well resolved experimentally.

In Fig.~\ref{UGa3_band_FS} (d), we present a comparison of the Fermi surface map and the result of the band-structure calculation, which is the same as that in Fig.~\ref{UGa3_band_FS} (b) but within the \lineXXX{X}{R}{M} high-symmetry plane.
There is a reasonable agreement between the experimental and the calculated results.
The calculated Fermi surface within this plane consists mainly of band 26, which forms a large square-shaped Fermi surface around the \pnt{R} point.
This structure agrees with the experimental Fermi surface map.
The hole pocket formed by band 26 around the \Gm point also agrees with the experimental Fermi surface map.
Band 25 forms a tiny hole pocket, and it appears as spots with somewhat enhanced intensity midway along the \lineXX{X}{R} high-symmetry line.
There is a similar feature in the corresponding location of the experimental Fermi surface map, and there exists a similar Fermi surface in the experimental spectra.
To summarize, the topology of the Fermi surface of \UGa is also essentially explained by the band-structure calculation although the experimental band structure in the vicinity of \EF is renormalized due to the finite electron correlation effect.

\subsection{Antiferromagnetic transition in \UGa}\label{UGa3_AF}
To further understand the nature of the antiferromagnetic transition in \UGa, we present the comparison of ARPES spectra of \UGa measured in the paramagnetic and the antiferromagnetic phases.
Figure~\ref{UGa3_tmpdep} summarizes the ARPES spectra of \UGa measured in the paramagnetic phase ($75~\mathrm{K}$) and the antiferromagnetic phase ($20~\mathrm{K}$).
Figure~\ref{UGa3_tmpdep} (a) shows the Brillouin zones of \UGa in the paramagnetic and the antiferromagnetic phases.
The symmetry of the Brillouin zone is changed from simple cubic in the paramagnetic phase to face-centered cubic in the antiferromagnetic phase.
Note that the \pnt{M} and the \pnt{R} points in the paramagnetic Brillouin zone become equivalent to the \pnt{X} and the \Gm points in the antiferromagnetic Brillouin zone, respectively.
Correspondingly, the \lineXXX{R}{M}{R} and the \lineXXX{M}{X}{M} high-symmetry lines in the paramagnetic phase become equivalent to the \lineXXX{\Gamma}{X}{\Gamma} and the $\mathrm{X} - \mathrm{W}- \mathrm{X} - \mathrm{W}- \mathrm{X}$ high-symmetry lines in the antiferromagnetic phase, respectively.
Figures~\ref{UGa3_tmpdep}(b)--(d) show comparisons of the ARPES spectra measured in the paramagnetic and the antiferromagnetic phases along three high-symmetry lines.
The spectra are divided by the Fermi--Dirac function broadened by the instrumental energy resolution to avoid the influence of Fermi cut-off.
There are no recognizable changes in these spectra.
Especially, the \pnt{M} point in the paramagnetic phase becomes equivalent to the \pnt{X} point in the antiferromagnetic phase, but the enhanced intensity at the \pnt{M} point in the paramagnetic phase does not appears at the \pnt{X} point in the antiferromagnetic phase as shown in Fig.~\ref{UGa3_tmpdep} (d).
Generally, antiferromagnetic transition is observed as the emergence of back-folded replica bands owing to changes in the Brillouin zone, and formation of a hybridization gap at the crossing point \cite{Cr_ARPES1, Cr_ARPES2}.
By contrast, changes in spectral profiles due to an antiferromagnetic transition in a system with low $T_\mathrm{N}$ and small ordered moments are very small \cite{UN_ARPES}.
The absence of clear changes in the spectral profiles suggests that \UGa is a weak itinerant magnet, as in the case of $\mathrm{UN}$ \cite{UN_ARPES}.

\subsection{Discussion}
The \Uf states in \UAl and \UGa have a very itinerant nature in the ground state, suggesting that the antiferromagnetism of \UGa originates from itinerant \Uf electrons.
This is consistent with macroscopic properties of \UGa, such as its small ordered moment in the antiferromagnetic phase.
The experimentally obtained Fermi surfaces of \UAl and \UGa were explained reasonably well by the band-structure calculation.
By contrast, the band in the vicinity of \EF exhibits noticeable deviations from the results of the band-structure calculations, that is, bands are renormalized near \EF owing to the weak but finite electron correlation effect.
The core-level spectra of \UAl and \UGa also suggest the existence of finite electron correlation effects.
Therefore, the band-structure calculation is a reasonable starting point for describing the electronic structures of \UAl and \UGa, but the electron correlation effect must be considered for describing their electronic structures.
Furthermore, no significant temperature dependencies across $T_\mathrm{N}$ were observed in the ARPES spectra of \UGa; thus, \UGa can be considered as weak itinerant antiferromagnet.

An important question is the relationship between their electronic structures and magnetic properties in these compounds.
We have shown experimentally that the Fermi surfaces of \UAl and \UGa are very similar to each other, although there are a few minor differences in their details.
For example, the cubic hole-type Fermi surface centered at the \pnt{R} point was observed in both \UAl and \UGa, but the hole-type Fermi surface formed by band 25 and the electron-type Fermi surface formed by band 27 in \UGa were not observed in the case of \UAl. 
If the magnetism in \UGa  originates of such minor differences in their Fermi surfaces, these two Fermi surfaces might play an important role in the emergence of antiferromagnetism in \UGa. 
By contrast, the electron correlation effect might play an essential role in the magnetic properties of this series of compounds.
As seen in the core-level spectra of these compounds in Fig.~\ref{AIPES} (b), the electron correlation effect is enhanced in the order of \UAl to \UIn.
The N\'eel temperature is also enhanced in the same order if one assumes that the N\'eel temperature of \UAl is $T_\mathrm{N}<0~\mathrm{K}$. 
The slightly larger specific coefficient of \UGa than that of \UAl indicates that the density of states at \EF $N(E_\mathrm{F})$ are larger in \UGa than in \UAl, which might satisfy the Stoner criterion $IN(E_\mathrm{F}) > 1$ in \UGa where $I$ is the energy reduction due to the electron correlation.
Nevertheless, the very similar electronic structures of \UAl and \UGa suggest that they are located at the boundary between magnetic and non-magnetic states.
This is consistent with the weak itinerant magnetic nature of \UGa observed in the present study and the spin-fluctuation nature inferred by the resistivity and magnetic susceptibility measurements \cite{UAl3_dHvA}.
Notably, the rare-earth based compounds of the same crystal structure, which have very localized $4f$ states, also exhibit antiferromagnetic transitions.
Their Fermi surfaces have some similarities with these of \UAl and \UGa, although they have considerably more complex structures \cite{Onuki_review}.
Therefore, the mechanism of antiferromagnetic transition in actinide and rare-earth compounds might have different origins, even though both types of compounds exhibit antiferromagnetic transitions.

\section{CONCLUSION}
The electronic structures of \UAlGaIn were studied using photoelectron spectroscopy.
The valence band and the core-level spectra showed that the electron correlation effect increases in the order of \UAl to \UIn.
Especially, the core-level spectrum of \UIn is qualitatively different from those of \UAl and \UIn, suggesting that the electron correlation effect is strongly enhanced in \UIn.
The detailed band structures and the Fermi surfaces of \UAl and \UGa were clarified by ARPES, and their essential structures were explained by the band-structure calculations.
The  topologies of the Fermi surfaces of \UAl and \UGa are very similar, but there exist a few differences.
These differences or the electron correlation effect might play an essential role in their different magnetic properties.
No noticeable changes were observed in the ARPES spectra of \UGa across the antiferromagnetic transition, suggesting that the magnetism of \UGa is of the weak-itinerant type.

\acknowledgments
The authors thank D. Kaczorowski for stimulating discussion.
The experiment was performed under Proposal Numbers 2013A3820, 2014A3820, 2014B3820, 2015A3820, 2015B3820, and 2016A3810 at SPring-8 BL23SU.
The present work was financially supported by JSPS KAKENHI Grant Numbers 26400374 and 16H01084 (J-Physics).

\bibliographystyle{apsrev}
\bibliography{UGa3_UAl3}

\begin{thebibliography}{37}
\expandafter\ifx\csname natexlab\endcsname\relax\def\natexlab#1{#1}\fi
\expandafter\ifx\csname bibnamefont\endcsname\relax
  \def\bibnamefont#1{#1}\fi
\expandafter\ifx\csname bibfnamefont\endcsname\relax
  \def\bibfnamefont#1{#1}\fi
\expandafter\ifx\csname citenamefont\endcsname\relax
  \def\citenamefont#1{#1}\fi
\expandafter\ifx\csname url\endcsname\relax
  \def\url#1{\texttt{#1}}\fi
\expandafter\ifx\csname urlprefix\endcsname\relax\def\urlprefix{URL }\fi
\providecommand{\bibinfo}[2]{#2}
\providecommand{\eprint}[2][]{\url{#2}}

\bibitem[{\citenamefont{van Maaren et~al.}(1974)\citenamefont{van Maaren, van
  Daal, Buschow, and Schinkel}}]{UX3_SSC}
\bibinfo{author}{\bibfnamefont{M.}~\bibnamefont{van Maaren}},
  \bibinfo{author}{\bibfnamefont{H.}~\bibnamefont{van Daal}},
  \bibinfo{author}{\bibfnamefont{K.}~\bibnamefont{Buschow}}, \bibnamefont{and}
  \bibinfo{author}{\bibfnamefont{C.}~\bibnamefont{Schinkel}},
  \bibinfo{journal}{Solid State Commun.} \textbf{\bibinfo{volume}{14}},
  \bibinfo{pages}{145} (\bibinfo{year}{1974}).

\bibitem[{\citenamefont{Koelling et~al.}(1985)\citenamefont{Koelling, Dunlap,
  and Crabtree}}]{UX3}
\bibinfo{author}{\bibfnamefont{D.~D.} \bibnamefont{Koelling}},
  \bibinfo{author}{\bibfnamefont{B.~D.} \bibnamefont{Dunlap}},
  \bibnamefont{and} \bibinfo{author}{\bibfnamefont{G.~W.}
  \bibnamefont{Crabtree}}, \bibinfo{journal}{Phys. Rev. B}
  \textbf{\bibinfo{volume}{31}}, \bibinfo{pages}{4966} (\bibinfo{year}{1985}).

\bibitem[{\citenamefont{Cornelius et~al.}(1999)\citenamefont{Cornelius, Arko,
  Sarrao, Thompson, Hundley, Booth, Harrison, and Oppeneer}}]{UX3_dHvA}
\bibinfo{author}{\bibfnamefont{A.~L.} \bibnamefont{Cornelius}},
  \bibinfo{author}{\bibfnamefont{A.~J.} \bibnamefont{Arko}},
  \bibinfo{author}{\bibfnamefont{J.~L.} \bibnamefont{Sarrao}},
  \bibinfo{author}{\bibfnamefont{J.~D.} \bibnamefont{Thompson}},
  \bibinfo{author}{\bibfnamefont{M.~F.} \bibnamefont{Hundley}},
  \bibinfo{author}{\bibfnamefont{C.~H.} \bibnamefont{Booth}},
  \bibinfo{author}{\bibfnamefont{N.}~\bibnamefont{Harrison}}, \bibnamefont{and}
  \bibinfo{author}{\bibfnamefont{P.~M.} \bibnamefont{Oppeneer}},
  \bibinfo{journal}{Phys. Rev. B} \textbf{\bibinfo{volume}{59}},
  \bibinfo{pages}{14473} (\bibinfo{year}{1999}).

\bibitem[{\citenamefont{\=Onuki et~al.}(2004)\citenamefont{\=Onuki, Settai,
  Sugiyama, Takeuchi, Kobayashi, Haga, and Yamamoto}}]{Onuki_review_JPSJ}
\bibinfo{author}{\bibfnamefont{Y.}~\bibnamefont{\=Onuki}},
  \bibinfo{author}{\bibfnamefont{R.}~\bibnamefont{Settai}},
  \bibinfo{author}{\bibfnamefont{K.}~\bibnamefont{Sugiyama}},
  \bibinfo{author}{\bibfnamefont{T.}~\bibnamefont{Takeuchi}},
  \bibinfo{author}{\bibfnamefont{T.~C.} \bibnamefont{Kobayashi}},
  \bibinfo{author}{\bibfnamefont{Y.}~\bibnamefont{Haga}}, \bibnamefont{and}
  \bibinfo{author}{\bibfnamefont{E.}~\bibnamefont{Yamamoto}},
  \bibinfo{journal}{J. Phys. Soc. Jpn.} \textbf{\bibinfo{volume}{73}},
  \bibinfo{pages}{769} (\bibinfo{year}{2004}).

\bibitem[{\citenamefont{Sakai et~al.}(2009)\citenamefont{Sakai, Kambe,
  Tokunaga, Chudo, Tokiwa, Aoki, Haga, \ifmmode~\bar{O}\else \={O}\fi{}nuki,
  and Yasuoka}}]{UIn3_NMR}
\bibinfo{author}{\bibfnamefont{H.}~\bibnamefont{Sakai}},
  \bibinfo{author}{\bibfnamefont{S.}~\bibnamefont{Kambe}},
  \bibinfo{author}{\bibfnamefont{Y.}~\bibnamefont{Tokunaga}},
  \bibinfo{author}{\bibfnamefont{H.}~\bibnamefont{Chudo}},
  \bibinfo{author}{\bibfnamefont{Y.}~\bibnamefont{Tokiwa}},
  \bibinfo{author}{\bibfnamefont{D.}~\bibnamefont{Aoki}},
  \bibinfo{author}{\bibfnamefont{Y.}~\bibnamefont{Haga}},
  \bibinfo{author}{\bibfnamefont{Y.}~\bibnamefont{\ifmmode~\bar{O}\else
  \={O}\fi{}nuki}}, \bibnamefont{and}
  \bibinfo{author}{\bibfnamefont{H.}~\bibnamefont{Yasuoka}},
  \bibinfo{journal}{Phys. Rev. B} \textbf{\bibinfo{volume}{79}},
  \bibinfo{pages}{104426} (\bibinfo{year}{2009}).

\bibitem[{\citenamefont{Aoki et~al.}(2000)\citenamefont{Aoki, Watanabe, Inada,
  Settai, Sugiyama, Harima, Inoue, Kindo, Yamamoto, Haga, and
  \=Onuki}}]{UAl3_dHvA}
\bibinfo{author}{\bibfnamefont{D.}~\bibnamefont{Aoki}},
  \bibinfo{author}{\bibfnamefont{N.}~\bibnamefont{Watanabe}},
  \bibinfo{author}{\bibfnamefont{Y.}~\bibnamefont{Inada}},
  \bibinfo{author}{\bibfnamefont{R.}~\bibnamefont{Settai}},
  \bibinfo{author}{\bibfnamefont{K.}~\bibnamefont{Sugiyama}},
  \bibinfo{author}{\bibfnamefont{H.}~\bibnamefont{Harima}},
  \bibinfo{author}{\bibfnamefont{T.}~\bibnamefont{Inoue}},
  \bibinfo{author}{\bibfnamefont{K.}~\bibnamefont{Kindo}},
  \bibinfo{author}{\bibfnamefont{E.}~\bibnamefont{Yamamoto}},
  \bibinfo{author}{\bibfnamefont{Y.}~\bibnamefont{Haga}}, \bibnamefont{and}
  \bibinfo{author}{\bibfnamefont{Y.}~\bibnamefont{\=Onuki}},
  \bibinfo{journal}{J. Phys. Soc. Jpn.} \textbf{\bibinfo{volume}{69}},
  \bibinfo{pages}{2609} (\bibinfo{year}{2000}).

\bibitem[{\citenamefont{Kaczorowski}(2006)}]{UGa3_Kaczorowski}
\bibinfo{author}{\bibfnamefont{D.}~\bibnamefont{Kaczorowski}},
  \bibinfo{journal}{J. Phys. Soc. Jpn.} \textbf{\bibinfo{volume}{75}},
  \bibinfo{pages}{68} (\bibinfo{year}{2006}), \bibinfo{note}{and references
  therein}.

\bibitem[{\citenamefont{Tatetsu et~al.}(2011)\citenamefont{Tatetsu, Teruya,
  Shirono, and Maehira}}]{UAl3_Maehira}
\bibinfo{author}{\bibfnamefont{Y.}~\bibnamefont{Tatetsu}},
  \bibinfo{author}{\bibfnamefont{H.}~\bibnamefont{Teruya}},
  \bibinfo{author}{\bibfnamefont{S.}~\bibnamefont{Shirono}}, \bibnamefont{and}
  \bibinfo{author}{\bibfnamefont{T.}~\bibnamefont{Maehira}},
  \bibinfo{journal}{J. Phys. Soc. Jpn.} \textbf{\bibinfo{volume}{80}},
  \bibinfo{pages}{SA126} (\bibinfo{year}{2011}).

\bibitem[{\citenamefont{Murasik et~al.}(1974)\citenamefont{Murasik,
  Leciejewicz, Ligenza, and Zygmunt}}]{UGa3_AF}
\bibinfo{author}{\bibfnamefont{A.}~\bibnamefont{Murasik}},
  \bibinfo{author}{\bibfnamefont{J.}~\bibnamefont{Leciejewicz}},
  \bibinfo{author}{\bibfnamefont{S.}~\bibnamefont{Ligenza}}, \bibnamefont{and}
  \bibinfo{author}{\bibfnamefont{A.}~\bibnamefont{Zygmunt}},
  \bibinfo{journal}{Phys. Stat. Sol. (a)} \textbf{\bibinfo{volume}{23}},
  \bibinfo{pages}{K147} (\bibinfo{year}{1974}).

\bibitem[{\citenamefont{Kaczorowski et~al.}(1993)\citenamefont{Kaczorowski,
  Tro\ifmmode~\acute{c}\else \'{c}\fi{}, Badurski, B\"ohm, Shlyk, and
  Steglich}}]{UGaSn3}
\bibinfo{author}{\bibfnamefont{D.}~\bibnamefont{Kaczorowski}},
  \bibinfo{author}{\bibfnamefont{R.}~\bibnamefont{Tro\ifmmode~\acute{c}\else
  \'{c}\fi{}}}, \bibinfo{author}{\bibfnamefont{D.}~\bibnamefont{Badurski}},
  \bibinfo{author}{\bibfnamefont{A.}~\bibnamefont{B\"ohm}},
  \bibinfo{author}{\bibfnamefont{L.}~\bibnamefont{Shlyk}}, \bibnamefont{and}
  \bibinfo{author}{\bibfnamefont{F.}~\bibnamefont{Steglich}},
  \bibinfo{journal}{Phys. Rev. B} \textbf{\bibinfo{volume}{48}},
  \bibinfo{pages}{16425} (\bibinfo{year}{1993}).

\bibitem[{\citenamefont{Schoenes et~al.}(2000)\citenamefont{Schoenes, Barkow,
  Broschwitz, Oppeneer, Kaczorowski, and Czopnik}}]{UGa3_optical}
\bibinfo{author}{\bibfnamefont{J.}~\bibnamefont{Schoenes}},
  \bibinfo{author}{\bibfnamefont{U.}~\bibnamefont{Barkow}},
  \bibinfo{author}{\bibfnamefont{M.}~\bibnamefont{Broschwitz}},
  \bibinfo{author}{\bibfnamefont{P.~M.} \bibnamefont{Oppeneer}},
  \bibinfo{author}{\bibfnamefont{D.}~\bibnamefont{Kaczorowski}},
  \bibnamefont{and} \bibinfo{author}{\bibfnamefont{A.}~\bibnamefont{Czopnik}},
  \bibinfo{journal}{Phys. Rev. B} \textbf{\bibinfo{volume}{61}},
  \bibinfo{pages}{7415} (\bibinfo{year}{2000}).

\bibitem[{\citenamefont{Hiess et~al.}(2001)\citenamefont{Hiess, Boudarot, Coad,
  Brown, Burlet, Lander, Brooks, Kaczorowski, Czopnik, and Troc}}]{UGa3_Hiess}
\bibinfo{author}{\bibfnamefont{A.}~\bibnamefont{Hiess}},
  \bibinfo{author}{\bibfnamefont{F.}~\bibnamefont{Boudarot}},
  \bibinfo{author}{\bibfnamefont{S.}~\bibnamefont{Coad}},
  \bibinfo{author}{\bibfnamefont{P.~J.} \bibnamefont{Brown}},
  \bibinfo{author}{\bibfnamefont{P.}~\bibnamefont{Burlet}},
  \bibinfo{author}{\bibfnamefont{G.~H.} \bibnamefont{Lander}},
  \bibinfo{author}{\bibfnamefont{M.~S.~S.} \bibnamefont{Brooks}},
  \bibinfo{author}{\bibfnamefont{D.}~\bibnamefont{Kaczorowski}},
  \bibinfo{author}{\bibfnamefont{A.}~\bibnamefont{Czopnik}}, \bibnamefont{and}
  \bibinfo{author}{\bibfnamefont{R.}~\bibnamefont{Troc}},
  \bibinfo{journal}{Europhys. Lett.} \textbf{\bibinfo{volume}{55}},
  \bibinfo{pages}{267} (\bibinfo{year}{2001}).

\bibitem[{\citenamefont{Usuda et~al.}(2004)\citenamefont{Usuda, Igarashi, and
  Kodama}}]{UGa3_Usuda}
\bibinfo{author}{\bibfnamefont{M.}~\bibnamefont{Usuda}},
  \bibinfo{author}{\bibfnamefont{J.-i.} \bibnamefont{Igarashi}},
  \bibnamefont{and} \bibinfo{author}{\bibfnamefont{A.}~\bibnamefont{Kodama}},
  \bibinfo{journal}{Phys. Rev. B} \textbf{\bibinfo{volume}{69}},
  \bibinfo{pages}{224402} (\bibinfo{year}{2004}).

\bibitem[{\citenamefont{Rusz et~al.}(2004)\citenamefont{Rusz, Biasini, and
  Czopnik}}]{UGa3_positron}
\bibinfo{author}{\bibfnamefont{J.}~\bibnamefont{Rusz}},
  \bibinfo{author}{\bibfnamefont{M.}~\bibnamefont{Biasini}}, \bibnamefont{and}
  \bibinfo{author}{\bibfnamefont{A.}~\bibnamefont{Czopnik}},
  \bibinfo{journal}{Phys. Rev. Lett.} \textbf{\bibinfo{volume}{93}},
  \bibinfo{pages}{156405} (\bibinfo{year}{2004}).

\bibitem[{\citenamefont{Dervenagas et~al.}(1999)\citenamefont{Dervenagas,
  Kaczorowski, Bourdarot, Burlet, Czopnik, and Lander}}]{UGa3_neutron}
\bibinfo{author}{\bibfnamefont{P.}~\bibnamefont{Dervenagas}},
  \bibinfo{author}{\bibfnamefont{D.}~\bibnamefont{Kaczorowski}},
  \bibinfo{author}{\bibfnamefont{F.}~\bibnamefont{Bourdarot}},
  \bibinfo{author}{\bibfnamefont{P.}~\bibnamefont{Burlet}},
  \bibinfo{author}{\bibfnamefont{A.}~\bibnamefont{Czopnik}}, \bibnamefont{and}
  \bibinfo{author}{\bibfnamefont{G.}~\bibnamefont{Lander}},
  \bibinfo{journal}{Physica B} \textbf{\bibinfo{volume}{269}},
  \bibinfo{pages}{368} (\bibinfo{year}{1999}).

\bibitem[{\citenamefont{Kambe et~al.}(2005)\citenamefont{Kambe, Walstedt,
  Sakai, Tokunaga, Matsuda, Haga, and \ifmmode~\bar{O}\else
  \={O}\fi{}nuki}}]{UGa3_NMR}
\bibinfo{author}{\bibfnamefont{S.}~\bibnamefont{Kambe}},
  \bibinfo{author}{\bibfnamefont{R.~E.} \bibnamefont{Walstedt}},
  \bibinfo{author}{\bibfnamefont{H.}~\bibnamefont{Sakai}},
  \bibinfo{author}{\bibfnamefont{Y.}~\bibnamefont{Tokunaga}},
  \bibinfo{author}{\bibfnamefont{T.~D.} \bibnamefont{Matsuda}},
  \bibinfo{author}{\bibfnamefont{Y.}~\bibnamefont{Haga}}, \bibnamefont{and}
  \bibinfo{author}{\bibfnamefont{Y.}~\bibnamefont{\ifmmode~\bar{O}\else
  \={O}\fi{}nuki}}, \bibinfo{journal}{Phys. Rev. B}
  \textbf{\bibinfo{volume}{72}}, \bibinfo{pages}{184437}
  (\bibinfo{year}{2005}).

\bibitem[{\citenamefont{Sanchez et~al.}(2000)\citenamefont{Sanchez, Vulliet,
  Abd-Elmeguid, and Kaczorowski}}]{UGa3_Mossbauer}
\bibinfo{author}{\bibfnamefont{J.~P.} \bibnamefont{Sanchez}},
  \bibinfo{author}{\bibfnamefont{P.}~\bibnamefont{Vulliet}},
  \bibinfo{author}{\bibfnamefont{M.~M.} \bibnamefont{Abd-Elmeguid}},
  \bibnamefont{and}
  \bibinfo{author}{\bibfnamefont{D.}~\bibnamefont{Kaczorowski}},
  \bibinfo{journal}{Phys. Rev. B} \textbf{\bibinfo{volume}{62}},
  \bibinfo{pages}{3839} (\bibinfo{year}{2000}).

\bibitem[{\citenamefont{Reihl et~al.}(1985)\citenamefont{Reihl, Domke, Kaindl,
  Kalkowski, Laubschat, Hulliger, and Schneider}}]{UGa3_RPES}
\bibinfo{author}{\bibfnamefont{B.}~\bibnamefont{Reihl}},
  \bibinfo{author}{\bibfnamefont{M.}~\bibnamefont{Domke}},
  \bibinfo{author}{\bibfnamefont{G.}~\bibnamefont{Kaindl}},
  \bibinfo{author}{\bibfnamefont{G.}~\bibnamefont{Kalkowski}},
  \bibinfo{author}{\bibfnamefont{C.}~\bibnamefont{Laubschat}},
  \bibinfo{author}{\bibfnamefont{F.}~\bibnamefont{Hulliger}}, \bibnamefont{and}
  \bibinfo{author}{\bibfnamefont{W.~D.} \bibnamefont{Schneider}},
  \bibinfo{journal}{Phys. Rev. B} \textbf{\bibinfo{volume}{32}},
  \bibinfo{pages}{3530} (\bibinfo{year}{1985}).

\bibitem[{\citenamefont{Aoki et~al.}(2001)\citenamefont{Aoki, Suzuki, Miyake,
  Inada, Settai, Sugiyama, Yamamoto, Haga, \=Onuki, Inoue, Kindo, Sugawara,
  Sato, and Yamagami}}]{UGa3_Aoki}
\bibinfo{author}{\bibfnamefont{D.}~\bibnamefont{Aoki}},
  \bibinfo{author}{\bibfnamefont{N.}~\bibnamefont{Suzuki}},
  \bibinfo{author}{\bibfnamefont{K.}~\bibnamefont{Miyake}},
  \bibinfo{author}{\bibfnamefont{Y.}~\bibnamefont{Inada}},
  \bibinfo{author}{\bibfnamefont{R.}~\bibnamefont{Settai}},
  \bibinfo{author}{\bibfnamefont{K.}~\bibnamefont{Sugiyama}},
  \bibinfo{author}{\bibfnamefont{E.}~\bibnamefont{Yamamoto}},
  \bibinfo{author}{\bibfnamefont{Y.}~\bibnamefont{Haga}},
  \bibinfo{author}{\bibfnamefont{Y.}~\bibnamefont{\=Onuki}},
  \bibinfo{author}{\bibfnamefont{T.}~\bibnamefont{Inoue}},
  \bibinfo{author}{\bibfnamefont{K.}~\bibnamefont{Kindo}},
  \bibinfo{author}{\bibfnamefont{H.}~\bibnamefont{Sugawara}},
  \bibinfo{author}{\bibfnamefont{H.}~\bibnamefont{Sato}}, \bibnamefont{and}
  \bibinfo{author}{\bibfnamefont{H.}~\bibnamefont{Yamagami}},
  \bibinfo{journal}{J. Phys. Soc. Jpn.} \textbf{\bibinfo{volume}{70}},
  \bibinfo{pages}{538} (\bibinfo{year}{2001}).

\bibitem[{\citenamefont{Mannix et~al.}(2001)\citenamefont{Mannix, Stunault,
  Bernhoeft, Paolasini, Lander, Vettier, de~Bergevin, Kaczorowski, and
  Czopnik}}]{UGa3_MXS}
\bibinfo{author}{\bibfnamefont{D.}~\bibnamefont{Mannix}},
  \bibinfo{author}{\bibfnamefont{A.}~\bibnamefont{Stunault}},
  \bibinfo{author}{\bibfnamefont{N.}~\bibnamefont{Bernhoeft}},
  \bibinfo{author}{\bibfnamefont{L.}~\bibnamefont{Paolasini}},
  \bibinfo{author}{\bibfnamefont{G.~H.} \bibnamefont{Lander}},
  \bibinfo{author}{\bibfnamefont{C.}~\bibnamefont{Vettier}},
  \bibinfo{author}{\bibfnamefont{F.}~\bibnamefont{de~Bergevin}},
  \bibinfo{author}{\bibfnamefont{D.}~\bibnamefont{Kaczorowski}},
  \bibnamefont{and} \bibinfo{author}{\bibfnamefont{A.}~\bibnamefont{Czopnik}},
  \bibinfo{journal}{Phys. Rev. Lett.} \textbf{\bibinfo{volume}{86}},
  \bibinfo{pages}{4128} (\bibinfo{year}{2001}).

\bibitem[{\citenamefont{Sarma et~al.}(1994)\citenamefont{Sarma, Krummacher,
  Gudat, Lin, Zhou, Crow, and Koelling}}]{UIn3_RPES}
\bibinfo{author}{\bibfnamefont{D.}~\bibnamefont{Sarma}},
  \bibinfo{author}{\bibfnamefont{S.}~\bibnamefont{Krummacher}},
  \bibinfo{author}{\bibfnamefont{W.}~\bibnamefont{Gudat}},
  \bibinfo{author}{\bibfnamefont{C.}~\bibnamefont{Lin}},
  \bibinfo{author}{\bibfnamefont{L.}~\bibnamefont{Zhou}},
  \bibinfo{author}{\bibfnamefont{J.}~\bibnamefont{Crow}}, \bibnamefont{and}
  \bibinfo{author}{\bibfnamefont{D.}~\bibnamefont{Koelling}},
  \bibinfo{journal}{Physica B} \textbf{\bibinfo{volume}{199 -- 200}},
  \bibinfo{pages}{622} (\bibinfo{year}{1994}).

\bibitem[{\citenamefont{Tokiwa et~al.}(2001)\citenamefont{Tokiwa, Aoki, Haga,
  Yamamoto, Ikeda, Settai, Nakamura, and \=Onuki}}]{UIn3_dHvA}
\bibinfo{author}{\bibfnamefont{Y.}~\bibnamefont{Tokiwa}},
  \bibinfo{author}{\bibfnamefont{D.}~\bibnamefont{Aoki}},
  \bibinfo{author}{\bibfnamefont{Y.}~\bibnamefont{Haga}},
  \bibinfo{author}{\bibfnamefont{E.}~\bibnamefont{Yamamoto}},
  \bibinfo{author}{\bibfnamefont{S.}~\bibnamefont{Ikeda}},
  \bibinfo{author}{\bibfnamefont{R.}~\bibnamefont{Settai}},
  \bibinfo{author}{\bibfnamefont{A.}~\bibnamefont{Nakamura}}, \bibnamefont{and}
  \bibinfo{author}{\bibfnamefont{Y.}~\bibnamefont{\=Onuki}},
  \bibinfo{journal}{J. Phys. Soc. Jpn.} \textbf{\bibinfo{volume}{70}},
  \bibinfo{pages}{3326} (\bibinfo{year}{2001}).

\bibitem[{\citenamefont{Yokoya et~al.}(1998)\citenamefont{Yokoya, Sekiguchi,
  Saitoh, Okane, Nakatani, Shimada, Kobayashi, Takao, Teraoka, Hayashi, Sasaki,
  Miyahara, Harami, and Sasaki}}]{BL23SU}
\bibinfo{author}{\bibfnamefont{A.}~\bibnamefont{Yokoya}},
  \bibinfo{author}{\bibfnamefont{T.}~\bibnamefont{Sekiguchi}},
  \bibinfo{author}{\bibfnamefont{Y.}~\bibnamefont{Saitoh}},
  \bibinfo{author}{\bibfnamefont{T.}~\bibnamefont{Okane}},
  \bibinfo{author}{\bibfnamefont{T.}~\bibnamefont{Nakatani}},
  \bibinfo{author}{\bibfnamefont{T.}~\bibnamefont{Shimada}},
  \bibinfo{author}{\bibfnamefont{H.}~\bibnamefont{Kobayashi}},
  \bibinfo{author}{\bibfnamefont{M.}~\bibnamefont{Takao}},
  \bibinfo{author}{\bibfnamefont{Y.}~\bibnamefont{Teraoka}},
  \bibinfo{author}{\bibfnamefont{Y.}~\bibnamefont{Hayashi}},
  \bibinfo{author}{\bibfnamefont{S.}~\bibnamefont{Sasaki}},
  \bibinfo{author}{\bibfnamefont{Y.}~\bibnamefont{Miyahara}},
  \bibinfo{author}{\bibfnamefont{T.}~\bibnamefont{Harami}}, \bibnamefont{and}
  \bibinfo{author}{\bibfnamefont{T.~A.} \bibnamefont{Sasaki}},
  \bibinfo{journal}{J. Synchrotron Rad.} \textbf{\bibinfo{volume}{5}},
  \bibinfo{pages}{10} (\bibinfo{year}{1998}).

\bibitem[{\citenamefont{Saitoh et~al.}(2012)\citenamefont{Saitoh, Fukuda,
  Takeda, Yamagami, Takahashi, Asano, Hara, Shirasawa, Takeuchi, Tanaka, and
  Kitamura}}]{BL23SU2}
\bibinfo{author}{\bibfnamefont{Y.}~\bibnamefont{Saitoh}},
  \bibinfo{author}{\bibfnamefont{Y.}~\bibnamefont{Fukuda}},
  \bibinfo{author}{\bibfnamefont{Y.}~\bibnamefont{Takeda}},
  \bibinfo{author}{\bibfnamefont{H.}~\bibnamefont{Yamagami}},
  \bibinfo{author}{\bibfnamefont{S.}~\bibnamefont{Takahashi}},
  \bibinfo{author}{\bibfnamefont{Y.}~\bibnamefont{Asano}},
  \bibinfo{author}{\bibfnamefont{T.}~\bibnamefont{Hara}},
  \bibinfo{author}{\bibfnamefont{K.}~\bibnamefont{Shirasawa}},
  \bibinfo{author}{\bibfnamefont{M.}~\bibnamefont{Takeuchi}},
  \bibinfo{author}{\bibfnamefont{T.}~\bibnamefont{Tanaka}}, \bibnamefont{and}
  \bibinfo{author}{\bibfnamefont{H.}~\bibnamefont{Kitamura}},
  \bibinfo{journal}{J. Synchrotron Rad.} \textbf{\bibinfo{volume}{19}},
  \bibinfo{pages}{388} (\bibinfo{year}{2012}).

\bibitem[{\citenamefont{Fujimori et~al.}(2015)\citenamefont{Fujimori, Ohkochi,
  Kawasaki, Yasui, Takeda, Okane, Saitoh, Fujimori, Yamagami, Haga, Yamamoto,
  and \ifmmode~\bar{O}\else \={O}\fi{}nuki}}]{UGe2_UCoGe_ARPES}
\bibinfo{author}{\bibfnamefont{S.-i.} \bibnamefont{Fujimori}},
  \bibinfo{author}{\bibfnamefont{T.}~\bibnamefont{Ohkochi}},
  \bibinfo{author}{\bibfnamefont{I.}~\bibnamefont{Kawasaki}},
  \bibinfo{author}{\bibfnamefont{A.}~\bibnamefont{Yasui}},
  \bibinfo{author}{\bibfnamefont{Y.}~\bibnamefont{Takeda}},
  \bibinfo{author}{\bibfnamefont{T.}~\bibnamefont{Okane}},
  \bibinfo{author}{\bibfnamefont{Y.}~\bibnamefont{Saitoh}},
  \bibinfo{author}{\bibfnamefont{A.}~\bibnamefont{Fujimori}},
  \bibinfo{author}{\bibfnamefont{H.}~\bibnamefont{Yamagami}},
  \bibinfo{author}{\bibfnamefont{Y.}~\bibnamefont{Haga}},
  \bibinfo{author}{\bibfnamefont{E.}~\bibnamefont{Yamamoto}}, \bibnamefont{and}
  \bibinfo{author}{\bibfnamefont{Y.}~\bibnamefont{\ifmmode~\bar{O}\else
  \={O}\fi{}nuki}}, \bibinfo{journal}{Phys. Rev. B}
  \textbf{\bibinfo{volume}{91}}, \bibinfo{pages}{174503}
  (\bibinfo{year}{2015}).

\bibitem[{\citenamefont{Yeh and Lindau}(1985)}]{Atomic}
\bibinfo{author}{\bibfnamefont{J.}~\bibnamefont{Yeh}} \bibnamefont{and}
  \bibinfo{author}{\bibfnamefont{I.}~\bibnamefont{Lindau}},
  \bibinfo{journal}{Atomic Data and Nuclear Data Tables}
  \textbf{\bibinfo{volume}{32}}, \bibinfo{pages}{1} (\bibinfo{year}{1985}).

\bibitem[{\citenamefont{Ohkochi et~al.}(2008)\citenamefont{Ohkochi, Fujimori,
  Yamagami, Okane, Saitoh, Fujimori, Haga, Yamamoto, and \ifmmode~\bar{O}\else
  \={O}\fi{}nuki}}]{UB2_ARPES}
\bibinfo{author}{\bibfnamefont{T.}~\bibnamefont{Ohkochi}},
  \bibinfo{author}{\bibfnamefont{S.-i.} \bibnamefont{Fujimori}},
  \bibinfo{author}{\bibfnamefont{H.}~\bibnamefont{Yamagami}},
  \bibinfo{author}{\bibfnamefont{T.}~\bibnamefont{Okane}},
  \bibinfo{author}{\bibfnamefont{Y.}~\bibnamefont{Saitoh}},
  \bibinfo{author}{\bibfnamefont{A.}~\bibnamefont{Fujimori}},
  \bibinfo{author}{\bibfnamefont{Y.}~\bibnamefont{Haga}},
  \bibinfo{author}{\bibfnamefont{E.}~\bibnamefont{Yamamoto}}, \bibnamefont{and}
  \bibinfo{author}{\bibfnamefont{Y.}~\bibnamefont{\ifmmode~\bar{O}\else
  \={O}\fi{}nuki}}, \bibinfo{journal}{Phys. Rev. B}
  \textbf{\bibinfo{volume}{78}}, \bibinfo{pages}{165110}
  (\bibinfo{year}{2008}).

\bibitem[{\citenamefont{Fujimori
  et~al.}(2012{\natexlab{a}})\citenamefont{Fujimori, Ohkochi, Okane, Saitoh,
  Fujimori, Yamagami, Haga, Yamamoto, and \ifmmode~\bar{O}\else
  \={O}\fi{}nuki}}]{UN_ARPES}
\bibinfo{author}{\bibfnamefont{S.-i.} \bibnamefont{Fujimori}},
  \bibinfo{author}{\bibfnamefont{T.}~\bibnamefont{Ohkochi}},
  \bibinfo{author}{\bibfnamefont{T.}~\bibnamefont{Okane}},
  \bibinfo{author}{\bibfnamefont{Y.}~\bibnamefont{Saitoh}},
  \bibinfo{author}{\bibfnamefont{A.}~\bibnamefont{Fujimori}},
  \bibinfo{author}{\bibfnamefont{H.}~\bibnamefont{Yamagami}},
  \bibinfo{author}{\bibfnamefont{Y.}~\bibnamefont{Haga}},
  \bibinfo{author}{\bibfnamefont{E.}~\bibnamefont{Yamamoto}}, \bibnamefont{and}
  \bibinfo{author}{\bibfnamefont{Y.}~\bibnamefont{\ifmmode~\bar{O}\else
  \={O}\fi{}nuki}}, \bibinfo{journal}{Phys. Rev. B}
  \textbf{\bibinfo{volume}{86}}, \bibinfo{pages}{235108}
  (\bibinfo{year}{2012}{\natexlab{a}}).

\bibitem[{\citenamefont{Laubschat et~al.}(1988)\citenamefont{Laubschat, Grentz,
  and Kaindl}}]{Laub_UBe13}
\bibinfo{author}{\bibfnamefont{C.}~\bibnamefont{Laubschat}},
  \bibinfo{author}{\bibfnamefont{W.}~\bibnamefont{Grentz}}, \bibnamefont{and}
  \bibinfo{author}{\bibfnamefont{G.}~\bibnamefont{Kaindl}},
  \bibinfo{journal}{Phys. Rev. B} \textbf{\bibinfo{volume}{37}},
  \bibinfo{pages}{8082} (\bibinfo{year}{1988}).

\bibitem[{\citenamefont{Fujimori et~al.}(2016)\citenamefont{Fujimori, Takeda,
  Okane, Saitoh, Fujimori, Yamagami, Haga, Yamamoto, and
  \=Onuki}}]{SF_review_JPSJ}
\bibinfo{author}{\bibfnamefont{S.-i.} \bibnamefont{Fujimori}},
  \bibinfo{author}{\bibfnamefont{Y.}~\bibnamefont{Takeda}},
  \bibinfo{author}{\bibfnamefont{T.}~\bibnamefont{Okane}},
  \bibinfo{author}{\bibfnamefont{Y.}~\bibnamefont{Saitoh}},
  \bibinfo{author}{\bibfnamefont{A.}~\bibnamefont{Fujimori}},
  \bibinfo{author}{\bibfnamefont{H.}~\bibnamefont{Yamagami}},
  \bibinfo{author}{\bibfnamefont{Y.}~\bibnamefont{Haga}},
  \bibinfo{author}{\bibfnamefont{E.}~\bibnamefont{Yamamoto}}, \bibnamefont{and}
  \bibinfo{author}{\bibfnamefont{Y.}~\bibnamefont{\=Onuki}},
  \bibinfo{journal}{J. Phys. Soc. Jpn.} \textbf{\bibinfo{volume}{85}},
  \bibinfo{pages}{062001} (\bibinfo{year}{2016}).

\bibitem[{\citenamefont{Fujimori
  et~al.}(2012{\natexlab{b}})\citenamefont{Fujimori, Ohkochi, Kawasaki, Yasui,
  Takeda, Okane, Saitoh, Fujimori, Yamagami, Haga, Yamamoto, Tokiwa, Ikeda,
  Sugai, Ohkuni, Kimura, and \=Onuki}}]{Ucore}
\bibinfo{author}{\bibfnamefont{S.-i.} \bibnamefont{Fujimori}},
  \bibinfo{author}{\bibfnamefont{T.}~\bibnamefont{Ohkochi}},
  \bibinfo{author}{\bibfnamefont{I.}~\bibnamefont{Kawasaki}},
  \bibinfo{author}{\bibfnamefont{A.}~\bibnamefont{Yasui}},
  \bibinfo{author}{\bibfnamefont{Y.}~\bibnamefont{Takeda}},
  \bibinfo{author}{\bibfnamefont{T.}~\bibnamefont{Okane}},
  \bibinfo{author}{\bibfnamefont{Y.}~\bibnamefont{Saitoh}},
  \bibinfo{author}{\bibfnamefont{A.}~\bibnamefont{Fujimori}},
  \bibinfo{author}{\bibfnamefont{H.}~\bibnamefont{Yamagami}},
  \bibinfo{author}{\bibfnamefont{Y.}~\bibnamefont{Haga}},
  \bibinfo{author}{\bibfnamefont{E.}~\bibnamefont{Yamamoto}},
  \bibinfo{author}{\bibfnamefont{Y.}~\bibnamefont{Tokiwa}},
  \bibinfo{author}{\bibfnamefont{S.}~\bibnamefont{Ikeda}},
  \bibinfo{author}{\bibfnamefont{T.}~\bibnamefont{Sugai}},
  \bibinfo{author}{\bibfnamefont{H.}~\bibnamefont{Ohkuni}},
  \bibinfo{author}{\bibfnamefont{N.}~\bibnamefont{Kimura}}, \bibnamefont{and}
  \bibinfo{author}{\bibfnamefont{Y.}~\bibnamefont{\=Onuki}},
  \bibinfo{journal}{J. Phys. Soc. Jpn.} \textbf{\bibinfo{volume}{81}},
  \bibinfo{pages}{014703} (\bibinfo{year}{2012}{\natexlab{b}}).

\bibitem[{\citenamefont{Schneider and Laubschat}(1981)}]{Laub_U4f}
\bibinfo{author}{\bibfnamefont{W.-D.} \bibnamefont{Schneider}}
  \bibnamefont{and}
  \bibinfo{author}{\bibfnamefont{C.}~\bibnamefont{Laubschat}},
  \bibinfo{journal}{Phys. Rev. Lett.} \textbf{\bibinfo{volume}{46}},
  \bibinfo{pages}{1023} (\bibinfo{year}{1981}).

\bibitem[{\citenamefont{Okada}(1999)}]{Okada_core}
\bibinfo{author}{\bibfnamefont{K.}~\bibnamefont{Okada}},
  \bibinfo{journal}{Journal of the Physical Society of Japan}
  \textbf{\bibinfo{volume}{68}}, \bibinfo{pages}{752} (\bibinfo{year}{1999}).

\bibitem[{\citenamefont{Zwicknagl}(2013)}]{Zwicknagl_core}
\bibinfo{author}{\bibfnamefont{G.}~\bibnamefont{Zwicknagl}},
  \bibinfo{journal}{Phys. Status Solidi B} \textbf{\bibinfo{volume}{250}},
  \bibinfo{pages}{634} (\bibinfo{year}{2013}).

\bibitem[{\citenamefont{Sch\"afer et~al.}(1999)\citenamefont{Sch\"afer,
  Rotenberg, Meigs, Kevan, Blaha, and H\"ufner}}]{Cr_ARPES1}
\bibinfo{author}{\bibfnamefont{J.}~\bibnamefont{Sch\"afer}},
  \bibinfo{author}{\bibfnamefont{E.}~\bibnamefont{Rotenberg}},
  \bibinfo{author}{\bibfnamefont{G.}~\bibnamefont{Meigs}},
  \bibinfo{author}{\bibfnamefont{S.~D.} \bibnamefont{Kevan}},
  \bibinfo{author}{\bibfnamefont{P.}~\bibnamefont{Blaha}}, \bibnamefont{and}
  \bibinfo{author}{\bibfnamefont{S.}~\bibnamefont{H\"ufner}},
  \bibinfo{journal}{Phys. Rev. Lett.} \textbf{\bibinfo{volume}{83}},
  \bibinfo{pages}{2069} (\bibinfo{year}{1999}).

\bibitem[{\citenamefont{Rotenberg et~al.}(2008)\citenamefont{Rotenberg, Krupin,
  and Kevan}}]{Cr_ARPES2}
\bibinfo{author}{\bibfnamefont{E.}~\bibnamefont{Rotenberg}},
  \bibinfo{author}{\bibfnamefont{O.}~\bibnamefont{Krupin}}, \bibnamefont{and}
  \bibinfo{author}{\bibfnamefont{S.~D.} \bibnamefont{Kevan}},
  \bibinfo{journal}{New J. Phy.} \textbf{\bibinfo{volume}{10}},
  \bibinfo{pages}{023003} (\bibinfo{year}{2008}).

\bibitem[{\citenamefont{\=Onuki and Settai}(2012)}]{Onuki_review}
\bibinfo{author}{\bibfnamefont{Y.}~\bibnamefont{\=Onuki}} \bibnamefont{and}
  \bibinfo{author}{\bibfnamefont{R.}~\bibnamefont{Settai}},
  \bibinfo{journal}{Low Temp. Phys.} \textbf{\bibinfo{volume}{38}},
  \bibinfo{pages}{89} (\bibinfo{year}{2012}).

\end{thebibliography}

\end{document}